\documentclass[lettersize,journal]{IEEEtran}
\usepackage{amsmath,amsfonts}
\usepackage{algorithmic}
\usepackage{algorithm}
\usepackage{array}
\usepackage[caption=false,font=normalsize,labelfont=sf,textfont=sf]{subfig}
\usepackage{textcomp}
\usepackage{stfloats}
\usepackage{url}
\usepackage{verbatim}
\usepackage{graphicx}
\usepackage{cite}
\usepackage{xspace}
\usepackage{scalerel}
\usepackage{amsmath, amssymb,mathptmx}

\newcommand{\jm}[1]{\mathbf{J}_{\mathrm{#1}}\xspace}
\newcommand{\pos}[1]{\mathrm{P}_{#1}\xspace}
\newcommand{\iti}{\mathit{i}}

\newcommand{\mean}[2]{\left< #1 \xspace\right>_{#2}\xspace}
\newcommand{\bmat}[1]{{\mathbf{#1}\xspace}}
\newcommand{\stretchedbar}[1]{\hstretch{1.67}{\bar{\hstretch{0.6}{#1}}}}
\newcommand{\etal}{\textit{et al.\@}\xspace}
\newcommand{\exvivo}{\textit{ex vivo}\xspace}
\newcommand{\Exvivo}{\textit{Ex vivo}\xspace}
\newcommand{\invivo}{\textit{in vivo}\xspace}

\newcommand{\invitro}{\textit{in vitro}\xspace}
\newcommand{\Invitro}{\textit{In vitro}\xspace}
\newcommand{\enface}{\textit{en face}\xspace}

\newcommand{\um}{{$\mu$m}\space}

\begin{document}

\title{Multi-contrast Jones-Matrix Optical Coherence Tomography --- The Concept, Principle, Implementation, and Applications}
\author{Yoshiaki Yasuno
\thanks{%
	This work was supported in part by the Japan Science and Technology Agency (JPMJCR2105, JPMJMI18G8) and Japan Society for Promotion of Science (21H01836, 22K04962, 21K09684, 18K09460).

	Y. Yasuno is with the Computational Optics Group, University of Tsukuba, Tsukuba, Ibaraki, Japan (e-mail: yoshiaki.yasuno@cog-labs.org).
			
	Color versions of one or more of the figures in this paper are available online at http://ieeexplore.ieee.org.

	Digital Object Identifier 10.1109/JSTQE.xxxx.xxxxxxx}}



\maketitle

\begin{abstract}
	Jones-matrix optical coherence tomography (JM-OCT) is an extension of polarization-sensitive OCT, which provides multiple types of optical contrasts of biological and clinical samples.
	JM-OCT measures the spatial distribution of the Jones matrix of the sample and also its time sequence.
	All contrasts (i.e., multi-contrast OCT images) are then computed from the Jones matrix.
	The contrasts obtained from the Jones matrix include not only the conventional and polarization-insensitive OCT intensity, cumulative and local phase retardation (birefringence), degree-of-polarization uniformity quantifying the polarization randomness of the sample, diattenuation, but also signal attenuation coefficient, sample scatterer density, Doppler OCT, OCT angiography, and dynamic OCT that contrasts intracellular motility or metabolism by analyzing the temporal fluctuation of the OCT signal.
	JM-OCT is a generalized version of OCT because it measures the generalized form of the sample information; i.e., the Jones matrix sequence.
	This review summarizes the basic conception, mathematical principle, hardware implementation, signal and image processing, and biological and clinical applications of JM-OCT.
	Advanced technical topics, including JM-OCT-specific noise correction and quantity estimation and JM-OCT's self-calibration nature, are also described.
\end{abstract}

\begin{IEEEkeywords}
Optical imaging, optical coherence tomography, polarization, optical coherence tomography angiography, three-dimensional microscopy
\end{IEEEkeywords}

\section{Introduction}
\subsection{Optical coherence tomography and Jones matrix optical coherence tomography}
\IEEEPARstart{O}{ptical} coherence tomography (OCT) \cite{Huang1991Science} is an optical measurement methodology that primarily measures the properties of a probe beam backscattered from a sample.
The properties of the sample are then obtained from the properties of the probe beam.
The probe beam can be characterized in terms of its amplitude, phase, and polarization.
Conventional OCT images are obtained from the amplitude of the probe beam, allowing visualization of the morphology of the sample.
Using the phase information of the beam, tiny structures and alterations of the probe path length smaller than the coherence-length-limited resolution can be measured \cite{Choma2005OL,Joo2005OL}.
The polarization of the probe beam is measured by hardware-extended OCT, polarization sensitive OCT (PS-OCT) \cite{Hee1992JOSAB, deBoer1997OL, deBoer2017BOE, Baumann2017AS}.
Through the probe-beam polarization, the polarization property of the sample can be measured.

Time sequential measurement can provide further information.
As an example, the flow of a fluid in a tissue can be measured by analyzing the temporal change in the phase of the probe beam.
This modality is known as Doppler OCT\cite{ZPChen1997b, Leitgeb2003Doppler, White2003OpEx, Makita2006OpEx, Leitgeb2014PRER, LYan2019TransBio}.
Similarly, tiny internal displacements in a sample can be measured from the phase change. 
Such measurements have been made to quantify the mechanical properties of tissues, such as
in optical coherence elastography \cite{Kennedy2014JSTQE,Larin2017BOE, Kennedy2021Book} and the assessment of laser-induced dynamic processes of tissues \cite{Muller2012BOE, Kurokawa2015BOE}.
Blood flow can also be visualized by measuring the statistical temporal fluctuation of the OCT signal amplitude, phase, or both.
This modality is referred to as OCT angiography (OCTA)\cite{Makita2006OpEx, RKWang2007BOE, CLChen2017BOE} and is widely used in the daily clinical routine of ophthalmology \cite{Spaide2018PRER}.
Slow dynamics of the tissue, such as those induced by tissue metabolism, have been assessed using similar methods, which is referred to as dynamic OCT (D-OCT)\cite{Apelian2016BOE, Munter2020OL, ElSadek2020BOE, Leung2020BOE, ElSadek2021BOE, Munter2021BOE}. 

Almost all properties of the probe beam, including its amplitude, phase, and polarization, can be represented using a Jones vector \cite{Jones1941JOSA}.
The time sequence of Jones vectors may contain the temporal properties of the probe beam.
Therefore, by measuring the sequence of Jones vectors, we can conduct not only PS-OCT but also Doppler OCT, OCTA, and D-OCT.

Although OCT directly measures the properties of the probe beam, the real interests of users are the properties of the tissue.
The properties of the probe beam are represented by Jones vectors, whereas the tissue properties are represented by the Jones matrix.
Jones-matrix OCT (JM-OCT) is a modality that primarily measures the Jones vectors, computes the Jones matrix, and gives multi-contrast images of the sample\cite{Yasuno2015Book_JmOct}.
The scattering and polarization properties of samples can be determined from the Jones matrix.
By sequentially measuring the Jones matrices, JM-OCT gives temporal properties of the sample, such as the flow and intracellular dynamics. 

Notably, the Jones vector and matrix are generalized representations of the probe beam and the sample's optical properties.
The time sequential Jones matrix is a more generalized sample representation because it represents the time dynamics of the sample in addition to the optical properties.
JM-OCT is thus the most generalized form of OCT.
This conception is discussed in more detail in Section \ref{sec:conception}.

\subsection{What is and is not covered by this review}
This review summarizes the basic conception (Section \ref{sec:conception}), principle (Section \ref{sec:principle}), implementation (Section \ref{sec:implementation}), and applications (Section \ref{sec:application}) of JM-OCT.
Sections \ref{sec:additionalTechnique} and \ref{sec:future} discuss additional technical topics and further extensions of JM-OCT.

It should be noted that, in this review, JM-OCT mainly refers to the modality developed at the University of Tsukuba, which is characterized by not only the Jones-matrix-based polarization measurement but also a multi-contrast imaging capability.
PS-OCT technology in its widest sense, including single-input-polarization PS-OCT\cite{Hee1992JOSAB, Hitzenberger2001OpEx, Pircher2004PMB, QZXiong2019OpEx, QZXiong2019BOE, PJTang2021Light}, Stokes-based PS-OCT\cite{deBoer1997OL}, and Jones--Muller-analysis-based PS-OCT\cite{Lippok2015OL, Villiger2016SciRep, Villiger2018Optica}, is not fully covered in this review and is only summarized briefly in Sections \ref{sec:PsOct} and \ref{sec:future}.
The reader can find good reviews of PS-OCT in the literature \cite{deBoer2017BOE, Baumann2017AS}.
One topic that is not thoroughly covered by these PS-OCT review papers is recently developed technologies for measuring the axis orientation  \cite{Villiger2018Optica, QYLi2018BOE, PJTang2021Light}.
This topic is briefly summarized in Section \ref{sec:axisOrientation} but extensive review of this topic is awaiting.

\section{Basic Conception of JM-OCT}
\label{sec:conception}
\begin{figure}
	\includegraphics[width=3.5in]{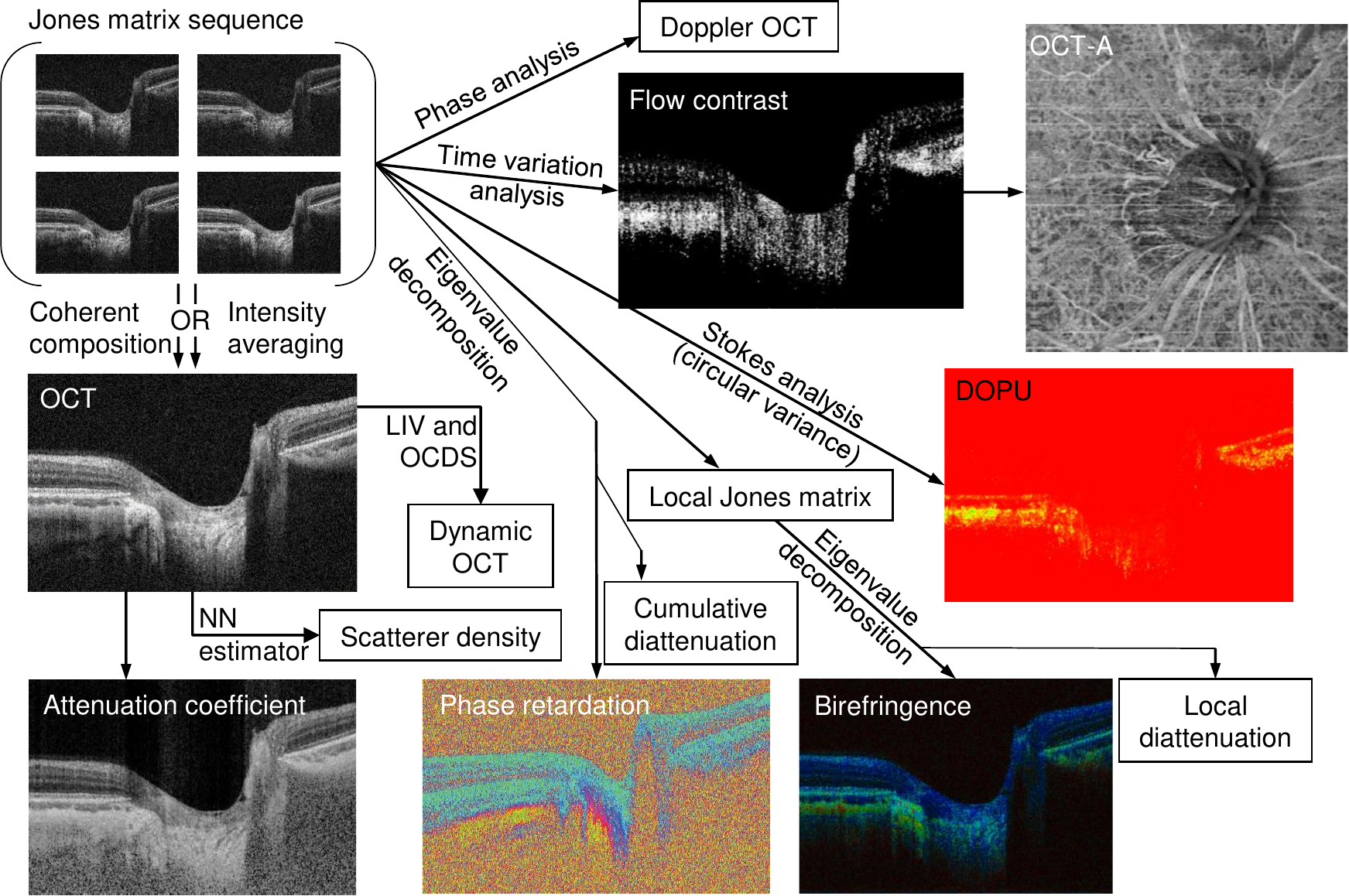}
	\caption{Basic concept of JM-OCT. 
		JM-OCT obtains the time sequence of Jones matrices of the sample.
		The Jones-matrix sequence comprises almost all the optical properties of the sample, which are extracted through proper signal processing.
		Intensity OCT images are obtained through coherent composition or intensity averaging.
		A polarization-artifact-free attenuation-coefficient image is obtained from the intensity average image.
		An attenuation-coefficient image and a scatterer-density image are obtained by processing the intensity OCT image.
		The latter is obtained using a neural-network-based scatterer density estimator.
		Two types of label-free tissue dynamics image (dynamic OCT; D-OCT) are obtained by processing the time sequence of the intensity images with two algorithms, namely the logarithmic-intensity variance (LIV) and OCT correlation decay speed (OCDS) algorithms.
		These algorithms  are sensitive to the magnitude and speed of the tissue dynamics, respectively.
		Through eigenvalue decomposition, cumulative-phase-retardation and cumulative-diattenuation images are obtained.
		A local Jones matrix is computed from the original Jones matrix, and birefringence and local-diattenuation images are obtained from the eigenvalues of the local Jones matrix.
		The degree-of-polarization uniformity (DOPU) is defined as a circular variance of Stokes vectors, which are obtained from the Jones matrix.
		Doppler OCT is obtained through temporal phase analysis and OCTA through temporal signal variation analysis.
	}
	\label{fig:conception}
\end{figure}
The basic conception of JM-OCT is presented in Fig.\@ \ref{fig:conception}.
JM-OCT measures a sequence of Jones matrices of the sample and regards it as the generalized representation of the sample.
Here, each of four entries of the Jones matrix is a complex OCT image.
In general, a Jones matrix is a complex 2-$\times$-2 matrix that transforms an incident Jones vector into an output Jones vector.
In principle, the four entries of the Jones matrix can thus be determined by measuring two output Jones vectors corresponding to two incident Jones vectors.

In JM-OCT, a polarization diversity (PD) detection, which splits the probe beam into two orthogonal polarization states and measures two interference signals, is adopted to measure the output Jones vector.
Here, each entry of the Jones vector is a complex OCT image corresponding to one of the two orthogonal polarizations of the backscattered probe beam.
In addition to the PD detection, two orthogonal polarization states are multiplexed in the incident probe beam by some means (see Section \ref{sec:implementation} for details).
JM-OCT thus measures two output Jones vectors corresponding to two incident polarization states; i.e., incident Jones vectors.
Each Jones vector comprises two complex OCT images, and four complex OCT images are thus obtained in a single measurement.
As detailed later in Section \ref{sec:principle}, these four OCT images form a ``measured Jones matrix ($\jm{m}$),'' which is mathematically similar to the Jones matrix of the sample.
By repeating this measurement, a time sequence of the Jones matrices is obtained.
This Jones-matrix sequence is used to compute several contrasts.



By applying a coherent composite method to the four entries of the Jones matrix, a high-sensitive OCT image is obtained (Section 3.6 of Ref.\@ \cite{MJJu2013OpEx}).
Intensity averaging of the two or four entries of the Jones matrix gives a polarization insensitive (i.e., birefringence-artifact-free) OCT image.
By processing this polarization sensitive OCT image, we also obtain an attenuation coefficient image \cite{Vermeer2014BOE}, which is a more direct representation of the tissue's scattering property than the raw OCT image.
The scatterer density image is a more direct measure of the tissue property than the attenuation coefficient and is obtained through a neural-network-based estimator.

The Jones-matrix images are obtained as a time sequence, and the intensity images can thus be obtained as a time sequence.
By processing the time-sequential OCT intensity image, two types of label-free tissue dynamics image (D-OCT) are obtained.

As the phase difference between two eigenvalues of the Jones matrix, a classical polarization sensitive OCT image (i.e., a cumulative phase retardation image) is obtained \cite{HylePark2004OL}.
By converting the Jones matrices into the local Jones matrix before computing the eigenvalues, a local phase retardation image, or equivalently a birefringence image, is obtained \cite{Makita2010OpEx}.
Notably, the conventional cumulative phase retardation represents the polarization properties of the backscattered probe beam, whereas the birefringence directly represents the polarization properties of the sample.
The birefringence image highlights fibrous tissues, including collagenous tissues, fibrosis, and nerve fibers.

In addition to the cumulative phase retardation and the birefringence, cumulative and local diattenuations can be computed from the cumulative and local Jones matrices.

The uniformity of the polarization can also be computed from the Jones matrix.
For example, the DOPU can be obtained through Stokes analysis of the Jones matrix\cite{Makita2014OL}.
The DOPU was originally introduced to non-Jones-matrix PS-OCT\cite{Gotzinger2008OpEx} and is sensitive to melanin\cite{Baumann2012BOE}.

A Doppler OCT image is obtained by analyzing the phase evolution of the Jones matrix entries or the global phase of the matrix \cite{MJJu2013OpEx, YJHong2014IOVS}.
In addition, an OCTA image can be obtained by quantifying the temporal fluctuation of the Jones matrix through correlation analysis \cite{Makita2016BOE}.

The details of the image generation of each contrast are described in the next section.

\section{Principle of JM-OCT}
\label{sec:principle}
\subsection{Cumulative Jones matrices and cumulative-phase-retardation measurement}
\label{sec:principle_cpr}
\begin{figure}
	\centering
	\includegraphics[width=2.5in]{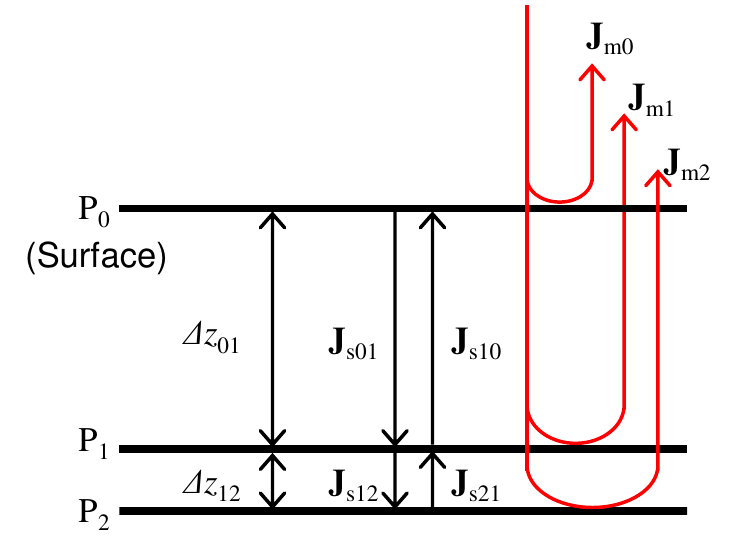}
	\caption{Model of the sample to be measured.
		Here the sample is modeled as having two layers with thicknesses of $\Delta z_{01}$ and  $\Delta z_{12}$.
		$\pos{0}$ to $\pos{2}$ are depth positions.
		$\jm{s\mathit{ij}}$ indicates the Jones matrix for a single trip from $\pos{i}$ to $\pos{j}$. 
		$\jm{m\mathit{i}}$ is the raw Jones matrix measured at $\pos{i}$.
		The figure is reprinted from \cite{Sugiyama2015BOE}.
	}
	\label{fig:sampleModel}
\end{figure}
The first step of the JM-OCT measurement is measuring a combined Jones matrix of the sample and the JM-OCT 	system.
Subsequently, a Jones matrix having the identical eigenvalues with the sample Jones matrix, i.e., a similar matrix of the sample Jones matrix, is computed.
To mathematically describe this process, we first model the sample as depicted in Fig.\@ \ref{fig:sampleModel}.
Here, the sample is modeled as having two layers.
The first layer is a superior layer from the tissue surface $\pos{0}$ to a depth $\pos{1}$ and has a thickness of $\Delta z_{01}$.
The second layer is a thin layer from $\pos{1}$ to $\pos{2}$ and has a thickness of $\Delta z_{12}$.
Note that the first layer can be thick and comprise multiple tissue layers whereas the second layer is assumed to be thin as it can be regarded as homogeneous along the depth.
The Jones matrices for forward and backward passes of a single trip between $\pos{i}$ and $\pos{j}$  are denoted $\jm{s\mathit{ij}}$ and $\jm{s\mathit{ji}}$, respectively.
The raw Jones matrices measured at $\pos{i}$ are denoted $\jm{m\mathit{i}}$.

Using Jones matrix hardware, which will be described later in Section \ref{sec:implementation}, $\jm{m\mathit{i}}$ is obtained as a function of space; i.e., an image of the Jones matrix.
The measured Jones matrix at the depth position $\pos{i}$ is expressed as
\begin{equation}
	\label{eq:measuredJM}
	\jm{m\mathit{i}} = \jm{out} \jm{r0\mathit{i}} \jm{in},
\end{equation}
where $\jm{out}$ and $\jm{in}$ are the Jones matrices of the collection and illumination optics, respectively.
These matrices are not a function of the space; i.e., they are constant matrices.
Hereafter, the two subscripts $i$ and $j$ of $\jm{r\mathit{ij}}$ indicate the indexes of depth positions, and $\jm{r\mathit{ij}}$ is the Jones matrix for a round trip from $\pos{i}$ to $\pos{j}$. 
That is to say, $\jm{r0\mathit{i}}$ is a round-trip sample Jones matrix from the surface ($\pos{0}$) to a depth position ($\pos{i}$), and $\jm{r0\mathit{i}} = \jm{s\mathit{i}0} \jm{s0\mathit{i}} = \jm{s0\mathit{i}}^T \jm{s0\mathit{i}}$.
Similar to $\jm{r\mathit{ij}}$, $\jm{s\mathit{ij}}$ is the Jones matrix for a single trip from $\pos{i}$ to $\pos{j}$.
Here, we use the fact that $\jm{s\mathit{i}0} = \jm{s0\mathit{i}}^T$.
The superscript $\mathrm{T}$ denotes a transpose.

In particular, the measured Jones matrix at the surface becomes a constant matrix
\begin{equation}
	\jm{m0} = \jm{out} \jm{in}.
\end{equation}

A cumulative Jones matrix at $\pos{\iti}$ is then computed as the product of the measured Jones matrix at $\pos{\iti}$ and the inverse of the surface Jones matrix
\begin{equation}
	\label{eq:cumulativeJm} 
	\jm{c\iti} \equiv \jm{m\iti} \jm{m0}^{-1} = \jm{out}\jm{r0\iti} \jm{out}^{-1},
\end{equation}
where $\jm{c\iti}$ is the cumulative Jones matrix.
As shown on the right-hand side of this equation, the cumulative Jones matrix is a similar matrix to the round-trip Jones matrix from the sample surface to $\pos{\iti}$, $\jm{r0\iti}$.
Owing to this similarity, the eigenvalues of the cumulative Jones matrix are identical to those of the round-trip sample Jones matrix.
The phase retardation, which is the phase difference between two eigenvalues of a Jones matrix, of this cumulative Jones matrix is thus identical to that of the round-trip sample Jones matrix.
Finally, a cumulative phase retardation, which is one of the polarization sensitive quantities measured by conventional PS-OCTs, is defined as the phase retardation computed from this cumulative Jones matrix.
Note that although the cumulative Jones matrix comprises the Jones matrices of the system optics ($\jm{out}$), the cumulative phase retardation solely relates to the sample Jones matrix because of the matrix similarity.

A more detailed formulation of the cumulative Jones matrix and cumulative phase retardation computation can be found in Sections 3.2 and 3.3 of Ref.\@ \cite{MJJu2013OpEx} and Section 2.2 of Ref.\@ \cite{Sugiyama2015BOE}.

\subsection{Local Jones matrices and local-phase-retardation measurement}
\label{sec:principle_lpr}
Although cumulative phase retardation has been widely used for PS-OCT imaging, it is not really interpretable unless the tissue between the tissue surface and the depth of interest has a uniform axis orientation.
Namely, if the axis orientation is not uniform, the eigenvalues of the cumulative Jones matrix cannot have a straight-forward relationship with those of the corresponding single-trip sample Jones matrix.
It makes the cumulative phase retardation uninterpretable (see Section 2.2 of Ref.\@ \cite{Sugiyama2015BOE} for details).

To overcome this limitation and obtain a more depth-localized polarization property, a local Jones matrix analysis has been introduced \cite{Makita2010OpEx}.
Assume the second layer in Fig.\@ \ref{fig:sampleModel} (from $\pos{1}$ to $\pos{2}$) is the layer of interest.
The round trip local Jones matrix of this layer is defined and computed as
\begin{equation}
	\label{eq:localJones}
	\jm{L12} \equiv \jm{c2} \jm{c1}^{-1} =  \jm{m2} \jm{m1}^{-1}.
\end{equation}
As shown in this equation, the local Jones matrix is defined from the cumulative Jones matrices mathematically (center part of the equation), but it can be directly computed from the measured Jones matrices (right hand side of the equation).

The local Jones matrix is written in a decomposed form as
\begin{equation}
	\jm{L12} = \mathbf{A} \Lambda^2_{s12} \mathbf{A}^{-1},
\end{equation}
where $\mathbf{A}$ is a matrix defined from the Jones matrices of the collection optics and the sample.
$\Lambda_{s12}$ is an eigenvalue matrix of the single-trip Jones matrix $\jm{s12}$, which is the Jones matrix of the layer of our interest.
The eigenvalue matrix is a diagonal matrix whose diagonal entries are the eigenvalues of $\jm{s12}$.
$\jm{L12}$ is a matrix similar to $\Lambda_{s12}^2$ as shown in this equation, and therefore no matter the values of $\mathbf{A}$, the phase retardation obtained from the local Jones matrix $\jm{L12}$ is identical to that obtained from $\Lambda_{s12}^2$.
$\Lambda_{s12}$ is a diagonal matrix by definition, and this phase retardation is thus twice the phase retardation of $\Lambda_{s12}$ and is a round-trip phase retardation of the layer of interest.	
This round-trip phase retardation is referred to as the local phase retardation of the thin layer from $\pos{1}$ to $\pos{2}$.

If the birefringence and optic axis orientation are uniform between $\pos{i}$ and $\pos{j}$, the tissue birefringence of this layer, $b_{ij}$, is obtained as
\begin{equation}
	b_{ij} = \frac{\mathrm{LPR}_{ij}}{2 k_0 \Delta z_{ij}},
\end{equation} 
where $\mathrm{LPR}_{ij}$ is the local phase retardation obtained from $\Lambda_{s12}$, $k_0$ is the center wavenumber of the probe beam, and $\Delta z_{ij}$ is the thickness of the layer.
A more detailed formulation of the local Jones matrix and local phase retardation computation can be found in Section 2.3 of Ref.\@ \cite{Sugiyama2015BOE}.

Note that the correct computation of the birefringence requires the uniformity of the layer of interest.
This requirement is, however, weaker than that of the measurement of the cumulative phase retardation.
That is to say, the measurement of cumulative phase retardation requires uniformity in the layer between the tissue surface and the depth of interest, which is notably thicker than the ``layer of interest'' of the local phase retardation measurement.
A more detailed derivation and requirement (assumptions) for the cumulative phase retardation and the birefringence measurement can be found in Sections 2, 6.1, 6.3 and Appendix A of Ref.\@ \cite{Sugiyama2015BOE}.

\subsection{Diattenuation}
\label{sec:diattenuation}
Diattenuations can be computed from the eigenvalues of a Jones matrix \cite{Chipman1989OptEng, Lu1996JOSAA}.
In the case of JM-OCT, cumulative and local diattenuations can be computed from the cumulative Jones matrix [Eq.\@ (\ref{eq:cumulativeJm})] and local Jones matrix [Eq.\@ (\ref{eq:localJones})], respectively, as
\begin{equation}
	d \equiv \frac{\left|\left| \lambda_1   \right|^2 - \left| \lambda_2  \right|^2  \right|}
	{\left| \lambda_1   \right|^2 + \left| \lambda_2  \right|^2},
\end{equation}
where $d$ is the diattenuation, $\lambda_1$ and $\lambda_2$ are two eigenvalues of the cumulative or local Jones matrices \cite{Yasuno2015Book_JmOct, deBoer2017BOE}.

Note that the diattenuation of biological tissue is negligible in general \cite{vanBlokland1985JOSAA, Brink1988KPSAA, Todorovic2004OL, HylePark2004OL, Kemp2005OpEx}.
And hence, the examples of diattenuation imaging are not shown in Section \ref{sec:application} (Applications).

\subsection{DOPU}
DOPU is a quantity first introduced by G\"otzinger \etal and strongly related to depolarization \cite{Gotzinger2008OpEx}.
The retinal pigment exhibits strong depolarization \cite{Miura2008IOVS} and low DOPU \cite{Gotzinger2008OpEx, Miura2008IOVS, Baumann2015IOVS}.
The DOPU has therefore been used to visualize the retinal pigment epithelium and its abnormality \cite{Pircher2011PRER, Miura2019SciRep, Miura2019IOVS, Miura2021SciRep}.

The DOPU is defined as the variance of the Stokes vectors on a Poincar\'e sphere, where the Stokes vectors are those measured within a small region in the image (i.e., the spatial kernel) and is mathematically defined as
\begin{equation}
	\mathrm{DOPU} \equiv \sqrt{
		\stretchedbar{Q}^2 + \stretchedbar{U}^2 + \stretchedbar{V}^2
	}\Big/\stretchedbar{I},
\end{equation}
where $\left(\stretchedbar{I}, \stretchedbar{Q}, \stretchedbar{U}, \stretchedbar{V}\right)$ denotes the averaged Stokes vector within the kernel \cite{Gotzinger2008OpEx}.


In the case of JM-OCT, two Stokes vectors that correspond to two incident polarization states are acquired at each pixel.
The two Stokes vectors are combined to compute a single DOPU ($\mathrm{DOPU'}$) as
\begin{equation}
	\mathrm{DOPU'} \equiv
	\mean{\sqrt{
		\stretchedbar{Q}^2 + \stretchedbar{U}^2 + \stretchedbar{V}^2}}{p}
	\Big/ \mean{\stretchedbar{I}}{p},
\end{equation}
where $\mean{\quad}{p}$ represents averaging over two polarization states \cite{Makita2014OL}.
In our particular implementation, we use a further modified definition of the DOPU to correct for the effect of measurement noise.
The details of this noise-corrected DOPU have been described elsewhere \cite{Makita2014OL}.

\subsection{Intensity OCT, attenuation coefficient, and scatterer density}
By combining the four OCT signals corresponding to the four Jones matrix entries, OCT intensity images are obtained as described in Section \ref{sec:conception}.
The attenuation coefficient image is obtained by further processing the OCT intensity.
In our particular implementation, we use the method of Vermeer \etal \cite{Vermeer2014BOE}.

The attenuation coefficient \cite{SChang2019JBO} is sensitive to tissue scattering and absorption and is suitable for highlighting tissue abnormalities.
The attenuation coefficient is, however, affected by system and measurement parameters, such as the depth of focus, focus position, and aberrations.
Several methods of correcting such effects have been demonstrated \cite{Smith2015IEEETMI, Stefan2018BOE, Kuebler2021BOE}.

In our JM-OCT, we additionally use scatterer density imaging achieved by a neural-network-based scatterer density estimator \cite{Seesan2022BOE}.
This estimator is based on a convolutional neural network trained by using numerically generated OCT images.
The images in the training dataset are generated from numerical samples with various scatterer densities.
The OCT resolution also varies from image to image.
The finally obtained estimator is thus insensitive to the system resolution and depth position of the focus.
This scatterer density imaging is a good and robust alternative to attenuation coefficient imaging.

\subsection{Doppler OCT and OCTA}
In JM-OCT, the Doppler OCT (i.e., Doppler shift between two complex OCT signals acquired at two time points) is computed as
\begin{equation}
	\Delta \phi(\mathbf{r}; t + \Delta t/2) = \mathrm{Arg}\left[\stretchedbar{E}(\mathbf{r}; t + \Delta t) \stretchedbar{E}(\mathbf{r}; t)^*\right],
\end{equation}
where $\stretchedbar{E}$ is a complex OCT signal obtained from the measured Jones matrix using the coherent composition method (Section 3.6 of Ref.\@ \cite{MJJu2013OpEx}).
$\mathbf{r}$ represents the spatial position, $t$ is time, and $\Delta t$ is the time difference between two measurements.
The superscript $*$ denotes the complex conjugate.
That is to say, $\stretchedbar{E}(\mathbf{r}; t)$ and $\stretchedbar{E}(\mathbf{r}; t + \Delta t)$ are OCT signals obtained at the same location but at different times.
They are typically B-scans obtained at the same location or A-scans obtained at very close positions.
Further details of the JM-OCT-based Doppler OCT method are given in Section 3.7 of Ref.\@ \cite{MJJu2013OpEx}.

OCTA can also be computed from the Jones matrix.
In principle, almost all OCTA algorithms designed for conventional OCT are applicable to JM-OCT.
As the simplest example, we can compute the coherent composition of complex OCT signals or the intensity-averaged OCT signal from the Jones matrix.
One of the conventional OCTA algorithms can be applied to these complex or intensity signals to yield an OCTA signal.
The examples of these algorithms include intensity based methods, such as speckle variance method \cite{Mariampillai2008OL}, phase based methods such as Doppler OCTA \cite{Makita2006OpEx} and phase variance methods \cite{Vakoc2009NatMed, Schwartz2014Ophthalmol}, and complex methods such as optical micro-angiography \cite{An2008OpEx}.
By processing each entry independently, we can also use split-spectrum amplitude-decorrelation angiography method (SSADA)  \cite{Jia2012OpEx}.
We can also more effectively use the four entries of each Jones matrix.
In our particular implementation, we use a complex-correlation-based OCTA algorithm that is tailored to JM-OCT\cite{Makita2016BOE} and provides high-quality OCTA images.
Notably, although it is not used in our JM-OCT implementation, Gong \etal recently presented a Jones-matrix-specific speckle decorrelation method for OCTA \cite{PJGong2020JBP}.

\subsection{Dynamic OCT (D-OCT)}
\label{sec:doct}
D-OCT, similar to OCTA, visualizes the temporal dynamics of the sample.
In contrast to OCTA, D-OCT uses more OCT frames acquired with a longer time period and contrasts more subtle dynamics of the tissue, such as intracellular motility or metabolism.
JM-OCT is compatible with several existing D-OCT methods.

In our JM-OCT modality, two D-OCT algorithms are used.
One algorithm is LIV \cite{ElSadek2020BOE}, which is defined as the variance of dB-scaled OCT intensities over a long acquisition time, typically around 6 s.
Here, the OCT intensity is computed as the intensity average of four Jones matrix entries.

The other algorithm is OCDS \cite{ElSadek2020BOE}, which is defined as the slope of the autocorrelation curve of the dB-scaled OCT signal.
In our particular implementation, two kinds of OCDS are defined with different delay time ranges of the autocorrelation slope.
One is early OCDS, which is defined for a short delay time, such as [12.8, 64 ms] \cite{ElSadek2020BOE}, and is sensitive to a fast fluctuation of the signal, i.e., fast dynamics of the tissue.
The other is late OCDS, which is defied for a long delay-time, such as [64, 627.2 ms]\cite{ElSadek2020BOE} or [204.8, 1228.8 ms]\cite{ElSadek2021BOE}, and is sensitive to slow dynamics.

The LIV quantifies the magnitude of the signal fluctuation whereas the OCDS is sensitive to the speed of the fluctuation.
They are thus complementary to each other.

Similar to the case of OCTA, JM-OCT is compatible with several other dynamic OCT algorithms, such as time spectral analysis \cite{Apelian2016BOE, Munter2020OL, Munter2021BOE, Leung2020BOE}, correlation analysis \cite{Oldenburg2015Optica}, and Brownian-bridge analysis \cite{Scholler2019OpEx}.

\section{Implementation of JM-OCT}
\label{sec:implementation}
\subsection{Generalized hardware}
\label{sec:generalHw}
\begin{figure}
	\centering\includegraphics[width=2.5in]{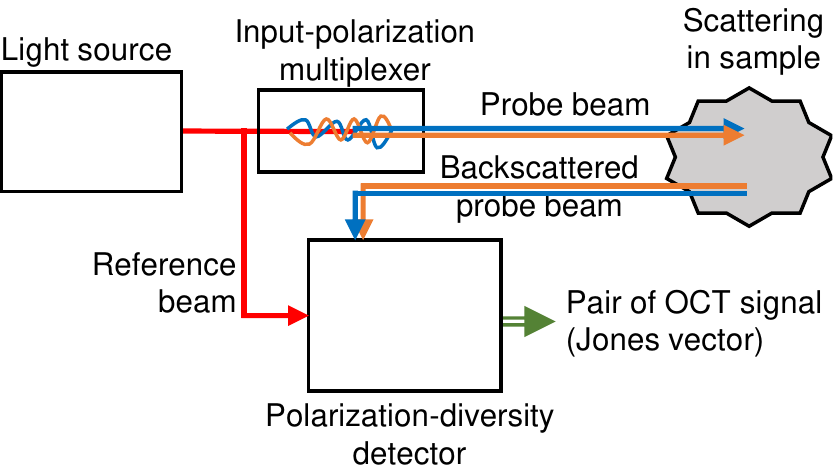}
	\caption{Generalized depiction of JM-OCT hardware.
		JM-OCT hardware is characterized by two polarization components: the input-polarization multiplexer and a PD detector.
		The input-polarization multiplexer multiplexes two polarization components of the incident probe beam, which are demultiplexed after OCT detection .
		The PD detector measures two interference signals that correspond to two output polarizations and it thus gives the Jones vector of the backscattered probe beam.
		\label{fig:generalHw}
	}
\end{figure}
As described in Section \ref{sec:conception}, JM-OCT uses two input polarization states and Jones vector detection.
Generalized hardware for such a detection scheme is depicted in Fig.\@ \ref{fig:generalHw}.
Here, two polarization-hardware components, namely the input-polarization multiplexer and PD detector module, are introduced.

The input-polarization multiplexer multiplexes two incident polarization states by some means as it can be demultiplexed after interference-signal detection.
For example, an electro-optic (EO) modulator can be used to modulate one of the polarization components with a particular frequency, so that two polarization states are multiplexed at different frequencies. 
In swept-source (SS)-OCT, the polarization components appear as two independent OCT images at two different depths\cite{Yamanari2008OpEx, Lim2011BOE}.
In spectral-domain (SD)-OCT, the EO modulator modulates the polarization states as a galvanometric mirror is performing a transversal scan (B-scan), so that the two polarization states are multiplexed into two different spatial frequencies along the transversal direction, and the states are demultiplexed later through numerical frequency filtering \cite{Yamanari2006OpEx}. 
As an alternative option, the EO modulator switches the incident polarization states for each A-line.
The two incident polarization components are then easily demultiplexed by de-interlacing the A-lines.
More modern swept-source JM-OCT adopts passive-delay-based incident-polarization multiplexing\cite{Baumann2012OpEx, Lim2012OL, Braaf2014BOE}.
Here, two different delays are applied to the two incident polarizations, and two OCT images corresponding to the two input polarizations thus appear at different depths in a single OCT image.
The polarization-dependent delay can be applied using bulk optics comprising polarization beam splitters \cite{Baumann2012OpEx, Lim2012OL, MJJu2013OpEx, Sugiyama2015BOE} or by a long polarization maintaining fiber that has different group delays between the two polarization states\cite{Baumann2012OpEx, ZWang2014BOE}.

The second polarization unit is a PD detector, which is used not only for JM-OCT but also for several types of polarization-sensitive \cite{Yamanari2006OpEx, Baumann2007OpEx, Cense2009OpEx, Braaf2014BOE, Yamanari2015BOE, Li2017BOE, Makita2018BOE, Baumann2018JBO, Azuma2019BOE, MJJu2019QIMS, QZXiong2019OpEx, PJTang2021Light} and non-polarization-sensitive OCT \cite{Vakoc2009NatMed, Makita2011OpEx, YSMiao2022OL}.
The PD detector splits the probe and reference beam, or equivalently the interference signal, into two polarization components and detects two interference signals corresponding to the two orthogonal polarizations of the backscattered probe beam using two photodetectors.
OCT can measure not only the amplitude but also the phase, and the PD detection scheme thus provides a pair of complex OCT signals corresponding to the Jones vector of the backscattered probe beam.

PD detection can be achieved using bulk \cite{Vakoc2009NatMed, MJJu2019QIMS} or semi-bulk optics \cite{Li2017BOE, Makita2018BOE, Azuma2019BOE} and also using polarization fiber components \cite{Yamanari2015BOE} or a combination of fiber and bulk optics \cite{Braaf2014BOE}.
In SS-OCT, the PD detector typically comprises two dual-balanced photodetectors.
In SD-OCT, the dual-camera spectrometer \cite{Yamanari2006OpEx, Makita2011OpEx} and/or dual spectrometer configuration \cite{Baumann2018JBO, QZXiong2019OpEx, PJTang2021Light, YSMiao2022OL} have been demonstrated.
Single-camera-based PD detection has also been demonstrated \cite{Baumann2007OpEx, Cense2009OpEx}.

The input-polarization multiplexer multiplexes two incident polarizations, the PD detector measures two output polarizations, and four OCT signals (i.e., the interactions of two inputs and two outputs) are thus acquired.
If both the input-polarization pair and output-polarization pair are orthogonal pairs, the four OCT signals evidently correspond to the four entries of the measured Jones matrix.
For example, if both the input-polarization and output-polarization pairs are the pairs of horizontal polarization ($[1\,\,\, 0]^\mathrm{T}$ in the Jones vector representation) and vertical polarization ($[0\,\, \,1]^\mathrm{T}$),
the measured Jones matrix is constructed as
\begin{equation}
	\jm{m} = \left[
	\begin{matrix}
		E_{HH} & E_{HV} \\
		E_{VH} & E_{VV}
	\end{matrix}
	\right],
	\label{eq:jmReconstruct}
\end{equation}
where $E$s are the complex OCT signals obtained with particular input and output polarization states.
The first and second subscripts indicate the input and output polarization states, respectively, where $H$ and $V$ denote horizontal and vertical.

Note that the complex OCT signals are the functions of the space, i.e., images.
And hence, the measured Jones matrix is also an image with four complex channels (i.e., the four matrix entries).
It is also noteworthy that the raw measured Jones matrix, $\jm{m\mathit{i}}$, in Eq.\@ (\ref{eq:measuredJM}) is the $\jm{m}$ of Eq.\@ (\ref{eq:jmReconstruct}) obtained at the surface $P_i$ in Fig.\@ \ref{fig:sampleModel}.

Even if the input- and output-polarization pairs are not orthogonal, we can use the four OCT signals as the four entries of the measured Jones matrix without modification.
The cumulative and local Jones matrix obtained from such measured Jones matrices is still similar to the matrix obtained with perfect orthogonality.  
Since a matrix and its similar matrix have identical eigenvalues, they give the same phase retardation values. 
The orthogonality is thus not a real requirement of JM-OCT.
The diattenuation and polarization-dependent phase offset of the system also do not affect the final results.
Even with these imperfections, the resulting local Jones matrix is again similar to the ideal one.
The details of this strong self-calibration nature of JM-OCT are given in Section \ref{sec:selfCalibration}.

\subsection{Implementation examples}
\subsubsection{Swept-source JM-OCT}
\begin{figure}
	\centering
	\includegraphics[width=3.5in]{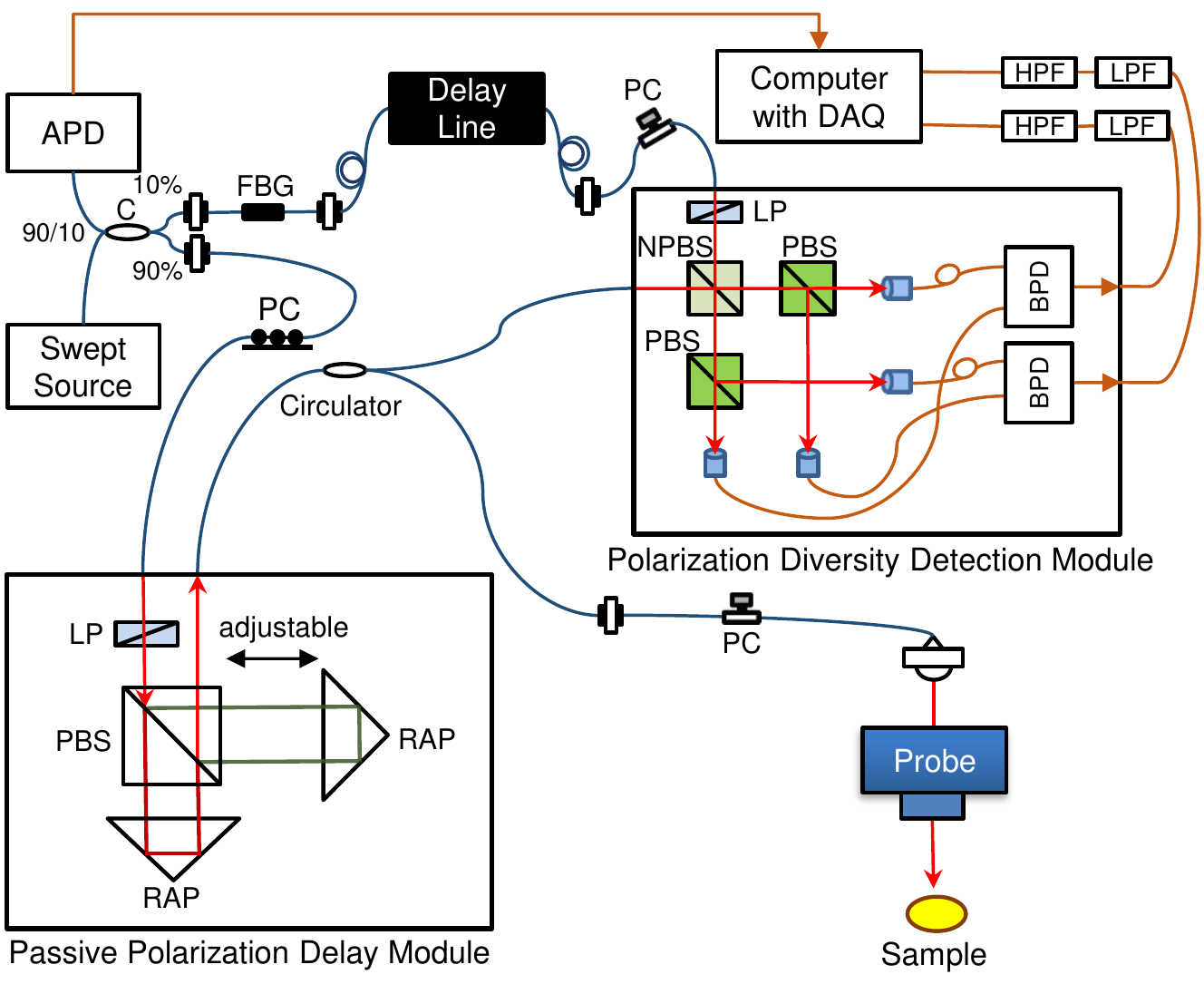}
	\caption{Implementation example of 1.3-\um swept-source Jones matrix OCT.
		The light from the light source (swept source) passes the passive-polarization-delay module before it is incident on the sample.
		The backscattered probe beam and the reference beam are introduced into the PD detection module, and two interference signals corresponding to two orthogonal polarizations are obtained.
		The data acquisition by the digitizer is triggered by a trigger generated by a fiber Bragg grating (FBG) and amplified photodetector (APD).
		C: circulator, RAP: right-angle prism, LP: linear polarizer, PC: polarization controller, PBS: polarization beam splitter, NPBS: non-polarizing beam splitter, HPF: high-pass filter, LPF: low-pass filter.
		The blue and brown lines indicate the optical fibers and electric cables, respectively.
	The figure is reprinted from Ref.\@ \cite{Li2017BOE} with modification.}
	\label{fig:transToadOptics}
\end{figure}
Figure \ref{fig:transToadOptics} is a schematic of 1.3-\um swept-source JM-OCT \cite{Li2017BOE}, which is shown as an implementation example.
Here, the light source is a microelectromechanical-system (MEMS) based wavelength sweeping laser source (Axsun Technologies, MA) with a center wavelength of 1,310 nm and a scanning rate of 49,600 Hz.
The light is split by a 90/10 fiber coupler (C), and the 10\% portion is sent to the reference delay line through a fiber Bragg grating (FBG, reflectivity $>$ 80\%).
The FBG sharply reflects a narrow spectrum light at 1,266 nm, and the reflected light is detected by an amplified photodetector (APD).
This triggers a digitizer for the acquisition of a spectral interference signal for each A-line.
Here, the digitizer is driven by a $k$-clock signal generated by the light source. 

The probe beam passes a semi-bulk, which is a well-encased small-optics, passive polarization delay module (Optohub Co. Ltd., Saitama, Japan) comprising a linear polarizer (LP), a polarization beam splitter (PBS), and two right-angle prisms (RAPs).
Here, the incident polarization is cleaned up in the 45-degree direction by the beam splitter and then split into p- and s-polarizations by the PBS.
Both polarized light beams are reflected by the RAPs, recombined by the PBS, and finally introduced into the probe optics through single-mode fibers and a circulator.
By displacing one of the RAPs, we arbitrarily set the mutual delay between the p- and s-polarized light beams.
Through this mutual polarization delay, two OCT images corresponding to these incident polarizations appear at two different depths in the imaging.
That is to say, two incident polarizations are multiplexed by the mutual delay and demultiplexed into the two different depths.

The interference signal is detected by a semi-bulk encased PD detection module (Optohub).
Here, the reference beam polarization is first cleaned up in the 45-degree direction.
After the reference and probe beams are mixed by the non-polarization beam splitter, the interference signal is split into horizontal and vertical polarization components by two PBSs and detected by two balanced photodetectors.
The interference signals are then cleaned up by low-pass and high-pass radio-frequency filters and digitized by a digitizer synchronized with the wavelength sweeping of the light source.
This PD diversity detection scheme simultaneously provides two interference signals, which correspond to two output polarizations.
Accounting for the input-polarization multiplexing by the passive polarization delay module, four OCT images are obtained at once, and they form the measured Jones matrix.
Multiple contrast images are obtained by processing this matrix using the method described in Section \ref{sec:principle}.

More details of this particular implementation can be found in \cite{Li2017BOE, Miyazawa2019BOE}.
Other variations but in a similar implementation of SS-JM-OCT can be found elsewhere \cite{Lim2012OL, Baumann2012OpEx, MJJu2013OpEx, Sugiyama2015BOE, XZhou2019JBO}.
It is also noteworthy that the passive polarization delay and the polarization diversity detection can be implemented not by using the bulk optics but using fiber components \cite{Baumann2012OpEx, ZWang2014BOE}.

\subsubsection{Spectral-domain JM-OCT}
\begin{figure}
	\centering
	\includegraphics[width=3.5in]{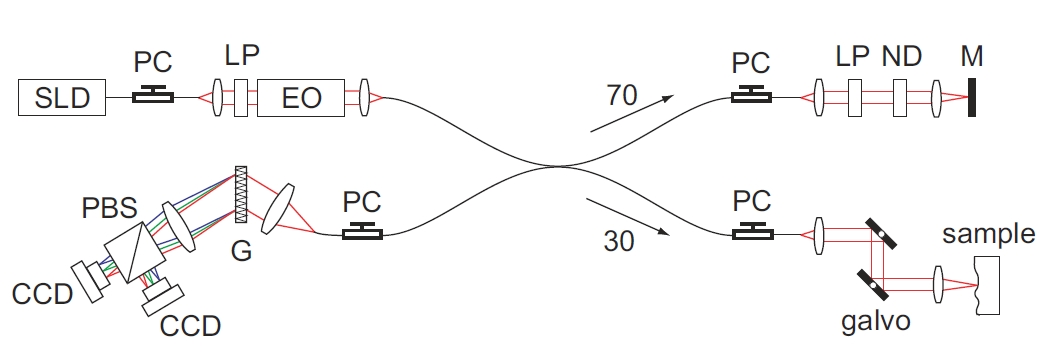}
	\caption{
		An implementation example of 840-nm SD-OCT-based JM-OCT.
		The light source is a superluminescent diode (SLD).
		The polarization state of the light is altered by an EO modulator (EO), which switches the light polarization for each A-line.
		The interference signal is detected by a PD spectrometer using a polarization beam splitter (PBS) and two line-scan cameras.
		G: grating, LP: linear polarizer, ND: neutral density filter, M, mirror, PC: polarization controller, galvo: galvanometric mirror. 
		The figure is reprinted from \cite{Yamanari2006OpEx}.
	}
	\label{fig:sdJmOctScheme}
\end{figure}
JM-OCT can be implemented also with SD-OCT technology as exemplified in Fig.\@ \ref{fig:sdJmOctScheme} (reprinted from Ref.\@ \cite{Yamanari2006OpEx}).
Here, an EO modulator modulates the input polarization along the transversal scan (B-scan), so that two incident polarization states can be demultiplexed by numerical frequency filtering along the transversal scan.
The PD detection is performed using a polarization-sensitive spectrometer in which two polarization components are split by a PBS and the interference signals corresponding to the two polarization components are detected by two line-scan cameras.
The combination of input-polarization switching and PD spectrometer provides four OCT signals with different input-output polarization combinations that form the measured Jones matrix.

In some PS-OCT, the polarization multiplexing is done with an acoustic optical modulator \cite{KHKim2011OpEx}.
The PD detection can also be achieved by several other configurations, such as a single-camera polarization sensitive spectrometer \cite{Baumann2007OpEx, Cense2009OpEx} and combinations of two identical spectrometers \cite{Baumann2018JBO, QZXiong2019OpEx, PJTang2021Light}.

Once the Jones matrix is measured, nearly identical signal processing with the SS-OCT-based JM-OCT can be conducted to yield images with multiple contrasts.

\section{Applications}
\label{sec:application}
JM-OCT has been applied to several clinical and biological fields, including investigation of the retina, anterior eye, skin, small animals, and cultured tissues.
Some of these applications are reviewed in this section.

\subsection{Retinal investigation}
\label{sec:retina}
The retina is one of the first and most successful application targets of JM-OCT. 
For clinical retinal investigation, several clinical prototype devices \cite{Sugiyama2015BOE, Azuma2019BOE}, including a function-limited simplified version \cite{Makita2018BOE}, have been developed.

Early studies targeted the quantification of retinal nerve fiber layers\cite{Yamanari2008JBO, Baumann2012OpEx, Braaf2014BOE} as a biomarker of glaucoma.
After PS-OCT studies showed that retinal pigments exhibit low DOPU\cite{Gotzinger2008OpEx}, JM-OCT was also adopted to investigate the retinal pigment epithelium (RPE) \cite{Baumann2012OpEx, Azuma2018BOE, Miura2019SciRep, Miura2019IOVS, Miura2021SciRep} and choroidal melanin \cite{Miura2017IOVS, Azuma2018BOE, Miura2022SciRep_VKH, Miura2022SciRep_ChMelanin} through DOPU imaging.
The former application was found to be useful in the investigation of exudative retinal diseases, such as age-related macular degeneration \cite{YJHong2014IOVS, Miura2017SciRep, Miura2021SciRep}, whereas the latter was found to be useful in the staging of Vogt--Koyanagi--Harada disease\cite{Miura2017IOVS, Miura2022SciRep_VKH}.

\begin{figure}
	\centering
	\includegraphics{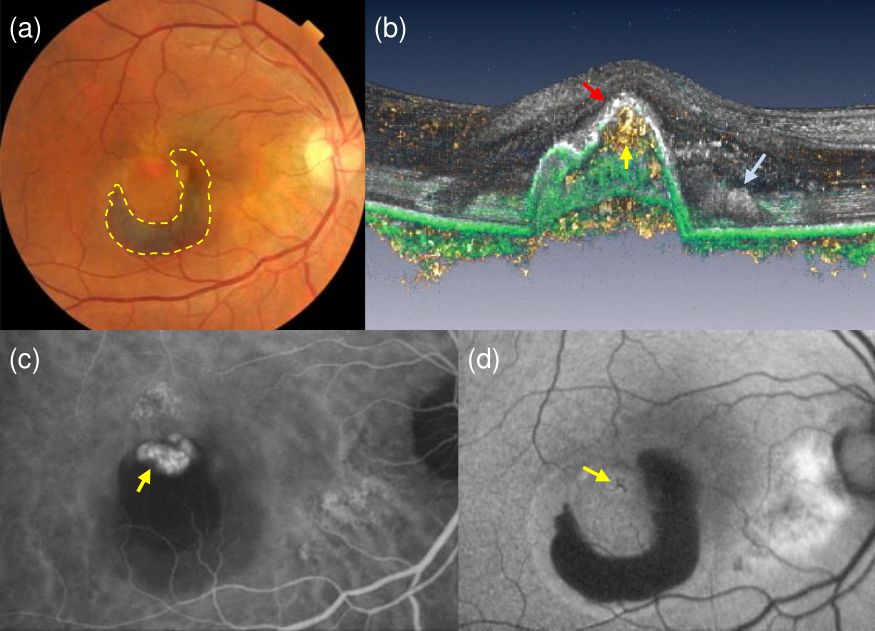}
	\caption{Example case of polypoidal choroidal vasculopathy:
		(a) color fundus photograph, (b) cut-away volume rendering of multi-contrast JM-OCT, (c) late-phase indocyanine green angiography, and FAF image.
		The color fundus (a) shows a large U-shaped hemorrhage region (dashed line). There is an active leakage in the indocyanine green angiography (c, arrow), and RPE damage is found in the same region in the FAF image (d).
		The corresponding multi-contrast JM-OCT image (b) shows all these clinical features.
		Gray, green, and yellow represent the OCT intensity, low-DOPU region (region with retinal pigments), and Doppler OCT signal (region with flow), respectively.
	}
	\label{fig:pcvMultiContrast}
\end{figure}
In addition to the sole use of polarization-sensitive imaging, its combination with flow imaging has been found to be useful in the investigation of exudative macular diseases\cite{YJHong2014IOVS}.
Figure \ref{fig:pcvMultiContrast} shows an example case of polypoidal choroidal vasculopathy (a new visualization of the case presented in Fig.\@ 7 of Ref.\@ \cite{YJHong2014IOVS}).
The color fundus (a) exhibits a U-shaped hemorrhage (delineated by a dashed line).
In the corresponding late-phase indocyanine green angiography (c), active leakage is found (arrow), and this leakage is collocated with the RPE defect found in the fundus auto-fluorescence (FAF) image (d, arrow).
These clinical features can be found in the three-dimensionally reconstructed multi-functional JM-OCT (b, cut-away volume reconstruction).
Here, gray represents the OCT intensity that corresponds to a conventional OCT signal.
Green represents low-DOPU regions, which indicate retinal pigments.
Yellow indicates flow acquired by Doppler OCT.
A large detachment of RPE is seen at the center of the image.
At the top of the RPE detachment (red arrow), a hyper-scattering layer corresponding to the RPE is shown, but this part of the RPE does not exhibit low-DOPU (green) signals.
This indicates that retinal pigments no longer exist in this region, which corresponds to the RPE damage shown in the FAF image (d, arrow).
A large cluster of flow (yellow arrow in b) is visible just beneath the RPE detachment.
This may be the abnormal vessel corresponding to the leakage found in the indocyanine green angiography (ICGA) image (arrow in c).
Next to the RPE detachment, a large hyper-scattering region without a flow signal is seen (blue arrow in b).
This region corresponds to the hemorrhage that appears in the color fundus (a).
No blood flow exists in this region as the hemorrhage is sufficiently old.  

For further assessment of retinal disease, several image processing methods tailored to multi-contrast JM-OCT have been demonstrated.
Azuma \etal arithmetically combined multiple contrasts to specifically highlight melanin associated with the RPE and choroid \cite{Azuma2018BOE}, which would be useful for the assessment of age-related macular degeneration \cite{Miura2019SciRep, Miura2021SciRep}.
Kasaragod \etal demonstrated a machine-learning-based method of quantitatively assessing the tissue properties of the lamina cribrosa \cite{Kasaragod2018BOE}, which is expected to be useful in glaucoma diagnosis.

Notably, several retinal JM-OCT implementations exist.
JM-OCT implementations were initially based on SD-OCT  \cite{Yamanari2008JBO}, but currently the majority are based on swept-source OCT with a 1-\um band probe beam and passive polarization multiplexing mechanism \cite{Lim2012OL, Baumann2012OpEx, MJJu2013OpEx, Braaf2014BOE, Sugiyama2015BOE} (see also Section \ref{sec:generalHw}).
In particular, the implementations of the authors started with a bulk laboratory prototype \cite{Yamanari2008JBO, Lim2012OL, MJJu2013OpEx}, and a compact clinical prototype based on the identical optical setup was demonstrated \cite{Sugiyama2015BOE}. 
By introducing semi-bulk packages of the polarization multiplexer and PD detector, a more compact and stable clinical system has been demonstrated \cite{Azuma2019BOE}.
A simplified retinal JM-OCT, which uses a PD detector but does not multiplex the input polarization, has also been demonstrated \cite{Makita2018BOE}.
Although this system cannot measure the phase retardation or birefringence, it can measure OCT, DOPU, and OCTA.
This simplified system is expected to enable low-cost clinical translation.
A motion-free and wide-field retinal scan method, namely the Lissajous scan method, has also been introduced into this system \cite{Makita2021BOE, Makita2022BOE}.
This system might enable comprehensive retinal investigation with low device cost.
Commercialization of retinal JM-OCT and its integration to a standard diagnostic flow are awaited.

\subsection{Anterior eye}
Another well-established target of JM-OCT is the anterior eye.
An early investigation was performed using a bulky laboratory prototype \cite{Yamanari2008OpEx,Yasuno2009OpEx}, and the first compact clinical prototype was then built by Lim \etal \cite{Lim2011BOE}.
This work has been followed by several more sophisticated clinical prototypes \cite{Yamanari2015BOE}.

The main targets of JM-OCT of the anterior eye are corneal diseases and glaucoma.
Among corneal diseases, JM-OCT is believed to be useful for the evaluation of keratoconus\cite{Fukuda2013IOVS}.
JM-OCT is expected to be used in the assessment of corneal collagen cross-linking, which is a keratoconus treatment \cite{MJJu2015JBO}.
The glaucoma applications of JM-OCT can be classified as risk assessments relating to anterior angle imaging and the follow up of glaucoma surgery. 
For the former, the polarization imaging of JM-OCT has been found to be useful in identifying the trabecular meshwork\cite{Yasuno2010JBO, Ueno2021TVST}.
For the latter, JM-OCT is used in investigating the fibrotic change in post-surgical structures of trabeculectomy \cite{Yasuno2009OpEx, Fukuda2014IOVS, Tsuda2015CXO, Kasaragod2016IOVS, Fukuda2016IOVS, Fukuda2018SciRep}.
In a JM-OCT based trabeculectomy study, Fukuda \etal pointed out that the phase retardation image can visualize slight fibrotic changes at early stage after the surgery, while the birefringence image can highlight localized fibrotic tissue \cite{Fukuda2018SciRep}.
In addition to these applications, the anterior sclera \cite{Yamanari2014BOE} and meibomian gland \cite{MJJu2014JBO} have been investigated by JM-OCT.

\subsection{Skin imaging}
Skin is a suitable target for JM-OCT because it comprises collagenous tissues.
Li \etal demonstrated the multi-contrast imaging of \invivo human skin adopting JM-OCT \cite{Li2017BOE}.
Kwon \etal demonstrated multi-contrast skin imaging \cite{SJKwon2017BOE}.
Notably, their device is specialized for skin assessment because it is equipped with a dermoscopic device .

Polarization, i.e., phase retardation and birefringence, investigations have revealed the age-related alternation of human skin \cite{Sakai2008JID}.
The relation between the polarization properties and structures of the skin surface, such as the wrinkle structure, has also been investigated \cite{Sakai2009JBO, Yamazaki2021SRT}.
The anisotropic response of dermal birefringence to mechanical stretching and compression has been found adopting JM-OCT\cite{Sakai2011BOE}.
Collagen is not only birefringent but also a strong source of second harmonic generation, and the combination of JM-OCT and second harmonic generation imaging is thus promising for the comprehensive investigation of skin \cite{VHLe2015BOE}.

In addition, the DOPU investigation \cite{XZhou2021BOE}, and ultra-high-resolution birefringence and optic axis imaging \cite{QYLi2020OL} of skin have recently been demonstrated.

\subsection{Zebrafish imaging}
Zebrafish are suitable for the close emulation of human diseases. Additionally, zebrafish have a short life span, providing a faster experimental cycle than other experimental animals.
JM-OCT has been adopted for the noninvasive and comprehensive investigation of zebrafish.

\begin{figure}
	\centering
	\includegraphics[width=3.5in]{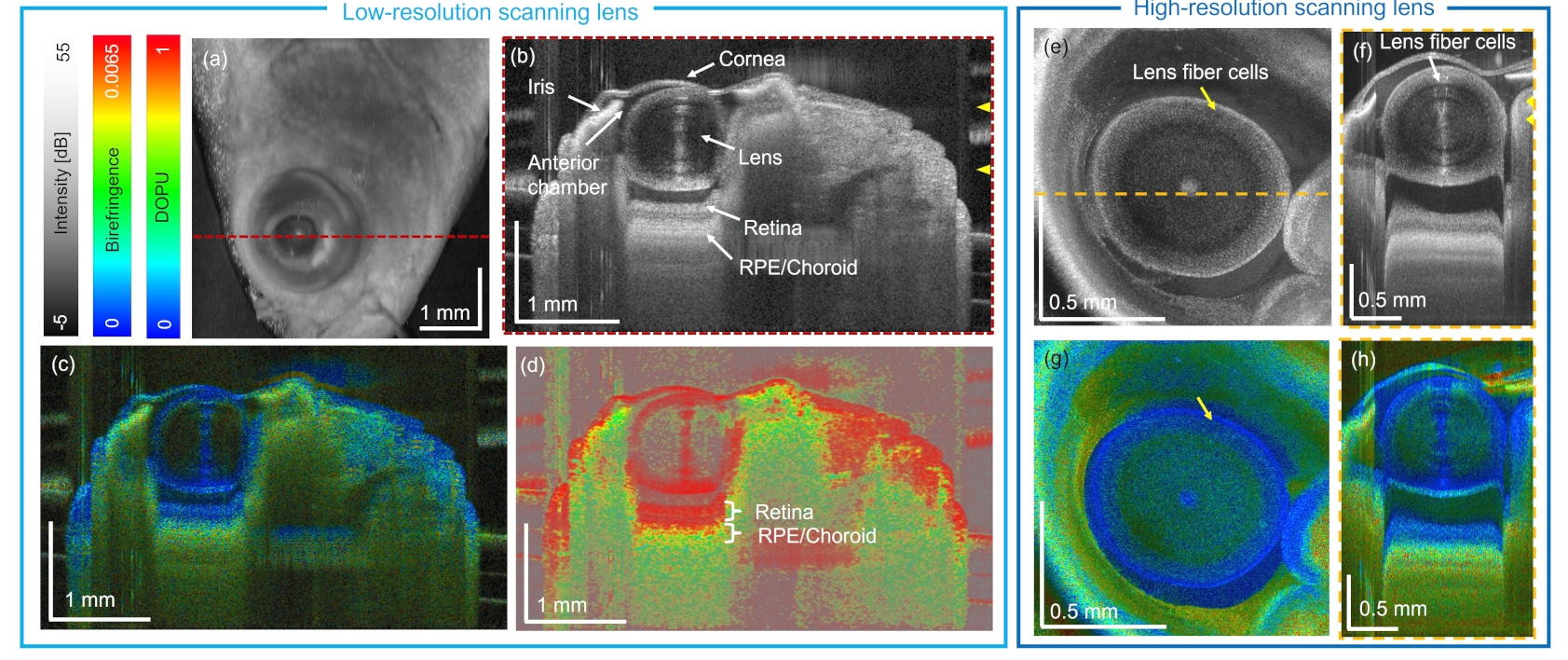}
	\caption{JM-OCT revealing fine  structures and polarization properties of a zebrafish eye.
		The images in the left box are the \enface OCT intensity projection (a) and intensity (b), birefringence (c), and DOPU (d) cross-sections obtained with a low-numerical-aperture objective with a lateral resolution of 18.1 \um.
		The high-resolution objective (right box, transversal resolution of 8.9 \um) reveals fine lens fiber structures from OCT intensity images (e, \enface projection and f, cross-section). 
		Corresponding birefringence images reveal low birefringence at the periphery of the lens.
		Dashed lines indicate the B-scan locations.
		The figure is reprinted from \cite{Lichtenegger2022BOE}.}
	\label{fig:lichtenegger2022BOE_fig2}
\end{figure}
\begin{figure}
	\centering
	\includegraphics[width=3.5in]{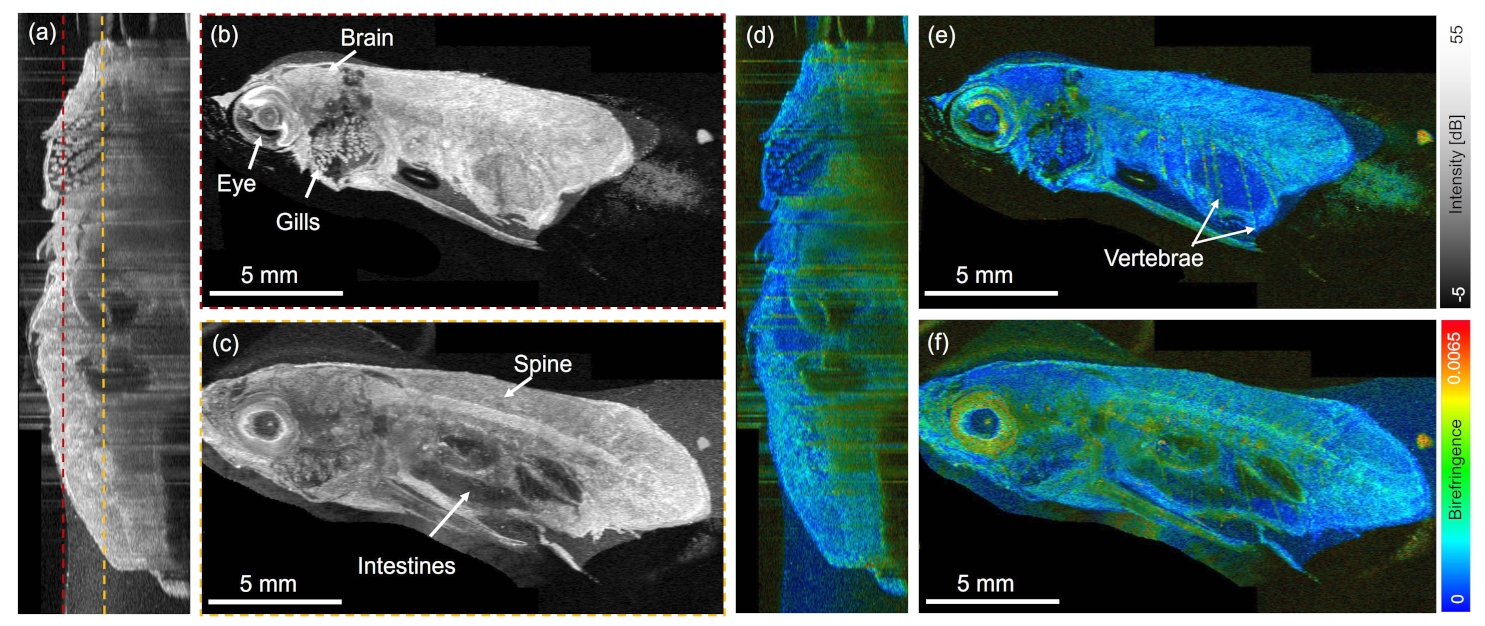}
	\caption{Whole-body zebrafish imaging by JM-OCT. Panels (a)--(c) show the OCT intensity and (d)--(f) the birefringence. 
		Panels (a) and (d) show results for the sagittal cross-section whereas panels (b), (c), (d), and (f) show the results for the \enface images.
		The \enface locations are indicated by red and orange dashed lines in (a).
		The figure is reprinted from \cite{Lichtenegger2022BOE}.
	}
	\label{fig:lichtenegger2022BOE_fig7}
\end{figure}
Lichtenegger \etal recently demonstrated several applications of JM-OCT to zebrafish.
The investigation of postmortem zebrafish revealed the fine structure of gills and eye globe \cite{Lichtenegger2022BOE}.
The birefringence and depolarization features of the eye globe were clearly visualized as shown in Fig.\@ \ref{fig:lichtenegger2022BOE_fig2} (reprinted from \cite{Lichtenegger2022BOE}).
Furthermore, by stitching multiple (three) JM-OCT volumes, the whole structure and birefringence of a zebrafish have been visualized as shown in Fig.\@ \ref{fig:lichtenegger2022BOE_fig7} (reprinted from \cite{Lichtenegger2022BOE}).

Juvenile and young zebrafish have been investigated \invivo through multi-contrast imaging, including the OCT intensity, birefringence, DOPU, and OCTA \cite{Lichtenegger2022JBO}.
The skeletal muscles and swim bladder were found to have high birefringence.
The RPE/choroid complex has a low DOPU, which may indicate melanin.
The quantitative analysis of birefringence revealed an increase in spinal birefringence with the growth of the fish (from 1 month to 2 months of age).

Human tumor cells can be xenotransplanted into zebrafish, and the zebrafish is thus expected to be an important sample for drug development.
The time evolution of human-derived tumor cells xenografted on zebrafish has also been shown to possibly be quantitatively evaluated through JM-OCT\cite{Lichtenegger2022SciRep}.
In the cited study, the temporal changes in the scattering, birefringence, and DOPU of a xenografted zebrafish were quantitatively investigated.
The OCTA images obtained by JM-OCT were also shown to be used to quantitatively evaluate the temporal changes  in the vasculature structure of the zebrafish.

\subsection{\Exvivo and \invitro tissue imaging}
\begin{figure}
	\centering
	\includegraphics[width=3.5in]{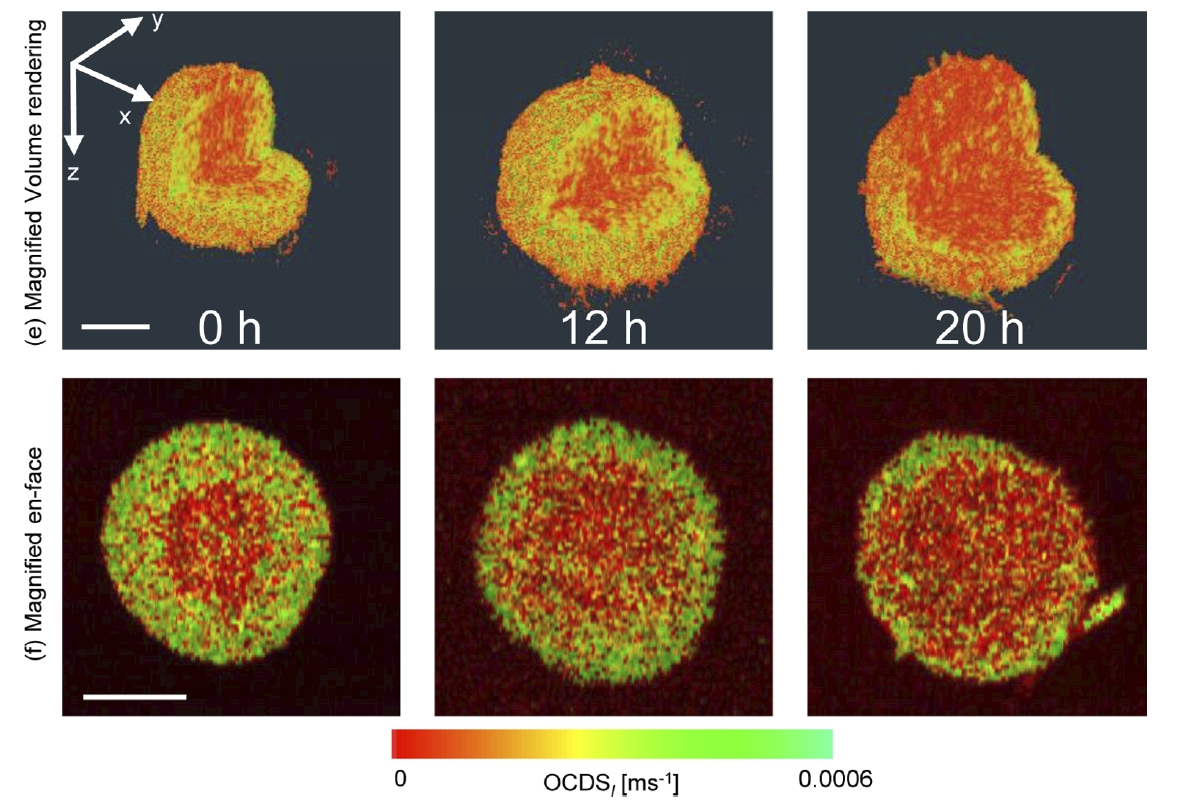}
	\caption{D-OCT of human breast cancer (MCF-7) spheroid visualized in a time course after extraction from the culturing environment.
		The first and second rows are three-dimensional volume renderings and \enface cross-sections, respectively.
		The center region shows a low D-OCT (OCDS$_l$) signal, which indicates necrotic cell death in the central region due to hypoxia and a lack of nutrients.
		The necrotic region becomes large over the time.
		The figures were adopted from \cite{ElSadek2021BOE}.
	}
	\label{fig:ElSadek2020BOE_fig4}
\end{figure}
\Invitro and \exvivo samples are recently emerging targets of JM-OCT because they are increasingly used for drug testing and development.
The temporal change and drug effect of tumor spheroids have been investigated\cite{ElSadek2020BOE, ElSadek2021BOE}.
Here, the D-OCT function of JM-OCT was used to visualize the cellular activities as exemplified in Fig.\@ \ref{fig:ElSadek2020BOE_fig4} (adopted from Ref.\@ \cite{ElSadek2021BOE}).
The temporal degradation of a human breast cancer (MFC-7) spheroid after extraction from a culturing environment was longitudinally and three-dimensionally visualized using the late OCDS (see Section \ref{sec:doct} for the details of the method).

\begin{figure}
	\centering
	\includegraphics[width=3.5in]{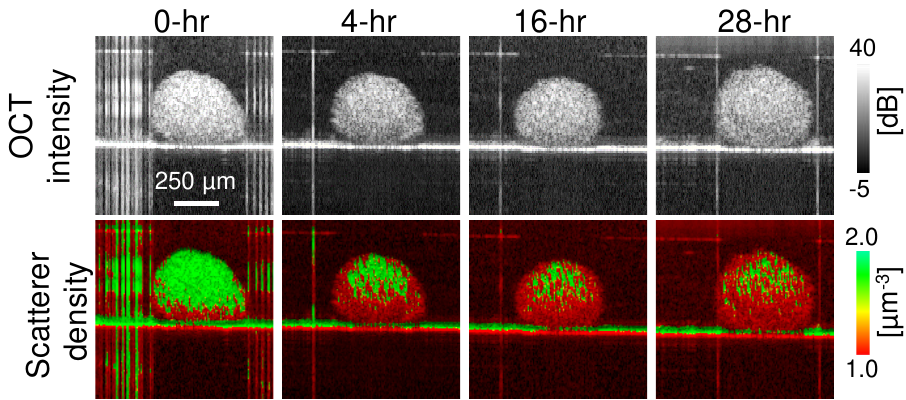}
	\caption{
		Scatterer density imaging revealing the temporal reduction in the scatterer density of a tumor spheroid.
		Hr denotes the time (hours) after the extraction of the spheroid from the culturing environment.
		The scatterer density reduces as the time elapses.
		The figure was adapted from Ref.\@ \cite{Seesan2022BOE} with modification.
	}
	\label{fig:spheroidSde}
\end{figure}
In addition to the D-OCT analysis, Seesan \etal recently applied neural-network-based scatterer-density analysis to the JM-OCT signal\cite{Seesan2022BOE}.
This revealed a temporal reduction in the scatterer density after the spheroid was extracted from the culturing environment as shown in Fig.\@ \ref{fig:spheroidSde}.
The multi-contrast capability of JM-OCT might enable the comprehensive investigation of \invitro tissues.

\begin{figure}
	\centering
	\includegraphics[width=3.5in]{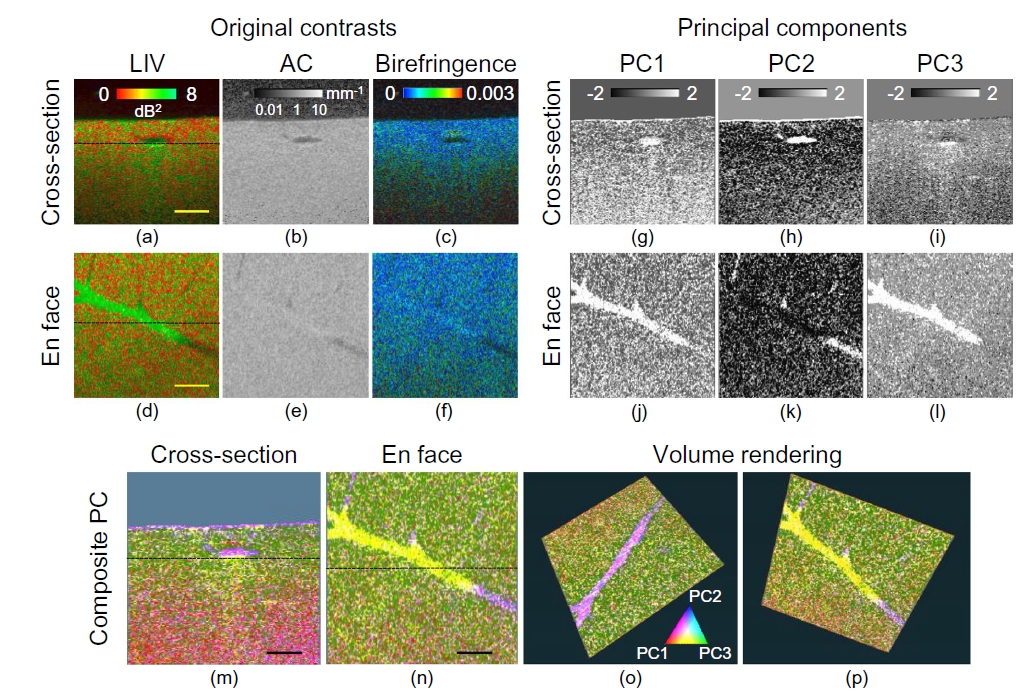}
	\caption{Multi-contrast imaging of mouse liver (a--f).
		The LIV image revealed fast dynamics (green) beneath the vessel stutterer even in the \exvivo sample.
		The LIV, attenuation coefficient (AC), and birefringence signals are converted into three principal component images (PC1 to PC3) in principal component analysis (g--l).
		By applying a tri-variate colormap (m--p), a blood vessel (purple) and active tissue beneath the vessel are clearly visualized. 
		The images were adopted from \cite{Mukherjee2021SciRep}.
	}
	\label{fig:Mukherjee2021SciRep_fig3}
\end{figure}
\begin{figure}
	\centering\includegraphics[width=3.5in]{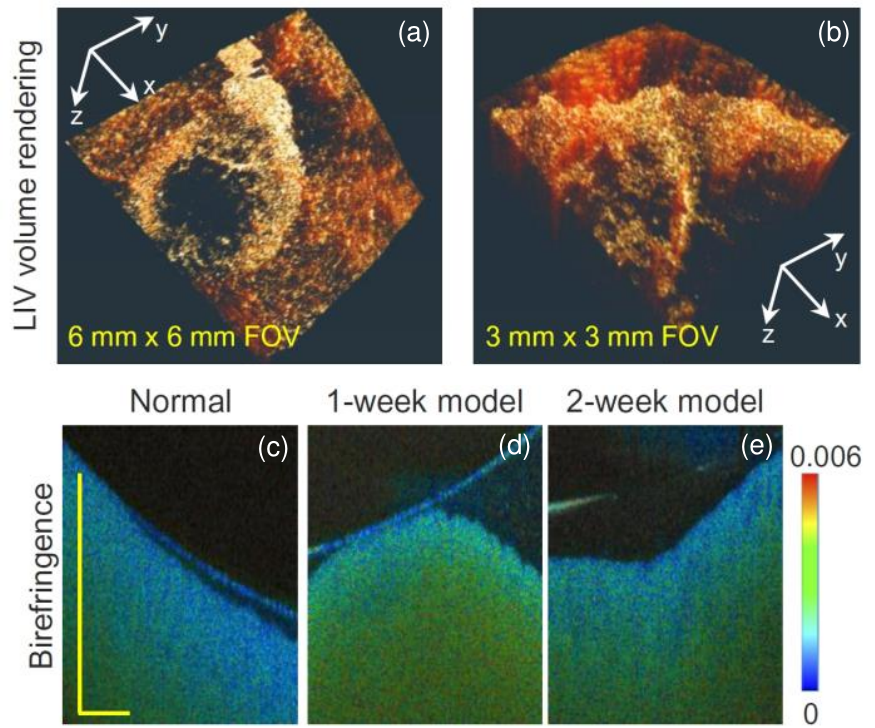}
	\caption{(a, b) Three-dimensional volume rendering of the LIV of 1-week NAFLD-model mouse liver.
		The imaging was performed with fields of view of 6 mm $\times$ 6 mm and 3 mm $\times$ 3 mm.
		Cross-sectional birefringence images of the normal mouse liver (c) and 1-week (d) and 2-week (e) NAFLD-model mouse livers.
		The 1-week model shows higher birefringence than the other livers.
		The images were reprinted with modification from \cite{Mukherjee2022BOE}.
	}
	\label{fig:Mukherjee2022BOE_fig2-4}
\end{figure}
\Exvivo mouse tissues have also been investigated adopting JM-OCT.
The structure and activities of the microvascular complex of mice have been visualized by combining simultaneously acquired attenuation coefficient, LIV, and birefringence images as shown in Fig.\@ \ref{fig:Mukherjee2021SciRep_fig3} \cite{Mukherjee2021SciRep}.
The D-OCT (LIV) investigation of normal and non-alcoholic-fatty-liver-disease (NAFLD) model mouse livers has revealed the high activity of lipid droplets, zonation structures, and inflammation that were not visible in conventional OCT imaging \cite{Mukherjee2022BOE}.
Here, the NAFLD models were 1-week and 2-week methionine choline-deficient (MCD)-diet models, which were fed with a MCD diet for 1 and 2 weeks.
Figure \ref{fig:Mukherjee2022BOE_fig2-4}(a, b) shows the three-dimensional distribution of lipid droplets of the 1-week NAFLD model.
Simultaneously obtained birefringence images show that the birefringence of the liver of the 1-week NAFLD model is higher than that of the normal liver, and it becomes low again in the liver of 2-week NAFLD model mouse as shown in Fig.\@ \ref{fig:Mukherjee2022BOE_fig2-4} (c-d) \cite{Mukherjee2022BOE}.  

\subsection{Further applications} 
JM-OCT has been adopted beyond the applications described above.
Tang and her colleagues have intensively investigated the polarization properties of cartilage \cite{XZhou2019JBO, XZhou2020BOE}.
Matsuzaki \etal adopted JM-OCT in assessing human-induced pluripotent-stem-cell-derived retinal pigment epithelial sheets using the polarization-sensitive function of JM-OCT \cite{Matsuzaki2020SciRep}.
Here, the measurement was made both \invitro and \invivo.
These applications suggest that JM-OCT can be used in regenerative medical procedures before and after transplantation.

\section{Additional technological topics}
\label{sec:additionalTechnique}
\subsection{Signal to noise ratio of JM-OCT}
The system sensitivity of OCT is known to be proportional to the probe beam power. Since the JM-OCT splits the probe power into four polarization channels, the sensitivity of each channel is four-times lower in average than the standard OCT with the same probe power.
This reduction of sensitivity can be compensated by coherently composite the four polarization channels (see Section 3.6 of Ref.\@ \cite{MJJu2013OpEx}).

It is also noteworthy that the signal-to-noise ratio (SNR) of the cumulative phase retardation and birefringence measurements is dominated by a quantity called ``effective SNR,’’ which is a reciprocal mean of SNRs of OCT signals involved in the computation of the cumulative phase retardation or birefringence (see Eq.\@ (39) of Ref.\@ \cite{Makita2010OpEx} for the mathematical definition).
In typical cases, the effective SNR is maximized when the probe power is equally distributed into the four polarization channels.
This gives us a simple optimization strategy of JM-OCT, i.e., as using a non-polarized mirror as a sample, adjusting the polarization controllers and polarizers in the system to have roughly equal signal strengths among the four polarization channels.

\subsection{Signal processing methods for better accuracy and image quality}
\label{sec:noiseCorrection}
Several signal processing techniques have been demonstrated to enhance the measurement accuracy and image quality of JM-OCT.
We listed examples of these technologies in this section.

To reduce the negative effect of noise, several signal enhancement and noise correction methods have been developed.
The first is adaptive Jones matrix averaging (AJA).
AJA is essentially complex averaging of the Jones matrix with a global phase correction.
The Jones matrix is a complex matrix, and the averaging of multiple Jones matrices thus suppresses the complex noise and fundamentally enhances the signal-to-noise ratio.
However, if the global phases fluctuate among the Jones matrices, the complex averaging decreases not only the noise but also the signal.
In AJA, the difference in the global phase between Jones matrices is estimated and corrected before the averaging.
AJA thus optimally enhances the SNR of the Jones matrix.
The details of the AJA method are described in Section 2.1.2 of Ref.\@ \cite{Lim2011BOE} and Section 3.4 of Ref.\@ \cite{MJJu2013OpEx}.

If the noise is additive and has a zero mean, such as in the case of complex noise in a complex OCT signal, the mean (averaging) of measured signals asymptotically approaches the true value.
However, the measured phase retardation and birefringence under noise are not symmetrically distributed around the true value \cite{Makita2010OpEx, Duan2011OpEx}.
This is because of the non-linear signal processing used to compute the phase retardation and birefringence from the set of complex OCT signals, i.e., the Jones matrix.
To accurately estimate the true birefringence from multiple measured birefringence values, maximum \textit{a-posteriori} estimators have been demonstrated\cite{Kasaragod2014OpEx, Kasaragod2017BOE}.
Here, Monte-Carlo simulation of JM-OCT imaging was used to compute the probability distribution of the measured birefringence under a particular true birefringence value and a particular effective SNR value.
Using a vast amount of simulation results with several configurations of the true birefringence and effective SNR, we can inversely determine the \textit{a-posteriori} distribution, which is the probability distribution of the true birefringence under particular values of the measured birefringence, and an effective SNR is acquired in measurements .
Using this estimator, a bias-free estimation of the birefringence can be obtained from the raw JM-OCT measurements.
The birefringence images shown in Figs.\@ \ref{fig:lichtenegger2022BOE_fig2}, \ref{fig:lichtenegger2022BOE_fig7}, \ref{fig:Mukherjee2021SciRep_fig3}, and \ref{fig:Mukherjee2022BOE_fig2-4} were obtained using this maximum \textit{a-posteriori} estimator.
A similar concept of maximum \textit{a-posteriori} estimation \cite{AChan2016OL} has been adopted to enhance the accuracy of the polarization-insensitive OCT intensity of JM-OCT \cite{AChan2017BOE}.

It should be noted that, for further improvement of the accuracy of phase retardation and birefringence measurement, it might be important to validate and optimize the measurement and estimation methods by using calibrated and standardized birefringent phantoms \cite{XYLiu2017BOE, QYLi2018BOE, PJTang2021Light, Chang2022JBO}.

The DOPU is also highly affected by noise.
The DOPU is the polarization uniformity among multiple measurements, and it is thus low if the SNR of the raw OCT signal is low.
Makita \etal presented a modified definition of the DOPU, which avoids the measurement-noise-originating reduction of the DOPU.
The details of this method are described in Ref.\@ \cite{Makita2014OL}.

In our JM-OCT, the OCTA signal is defined as temporal complex decorrelation of OCT signals, and it is thus also strongly affected by measurement noise.
To obtain high-contrast OCTA from the measured Jones matrices, a noise-immune OCTA method that estimates the decorrelation in correcting the phase noise effect, has been demonstrated \cite{Makita2016BOE}.

\subsection{Self-calibration nature of JM-OCT}
\label{sec:selfCalibration}
The JM-OCT theory described in Sections \ref{sec:principle_cpr} and \ref{sec:principle_lpr} assumes that both the multiplexed incident-polarization pair and the detected polarization pair are perfectly orthogonal pairs.
However, in reality, it is difficult to achieve such ideal orthogonality.
For example, a small amount of cross-talk in the PBS used in the passive polarization delay module and the PD detector degrades the orthogonality.
Alignment error of the PD detector causes a mutual phase shift between the two OCT signals of the two output polarizations, and it disturbs the ideal measurement of the Jones vector.
Although these imperfections are unavoidable in practical implementations, they are automatically canceled in the computation of cumulative and/or local Jones matrices. 
This self-calibration nature is further described as follows.

Considering the above factors, the measured Jones matrix at position $\pos{i}$ [Eq.\@ (\ref{eq:measuredJM})] is modified as 
\begin{equation}
	\jm{m\mathit{i}} = \eta\, \bmat{X'}\, \bmat{R}\, \rho\, \jm{out}\, \jm{r01\mathit{i}}\, \jm{in}\, \bmat{X}\, f\left(\bmat{X}\, \bmat{E}_{in}\,\right),
	\label{eq:JmImperfect}
\end{equation}
where $\bmat{E}_{in}$ is a matrix whose columns are two polarizations incident on the polarization delay module.
$\bmat{X}$ is a matrix representing the imperfection of the PBS in the polarization delay module (input-polarization multiplexer), and $f(\,)$ is a function representing the delay between two incident polarization states introduced by the delay module.
The off-diagonal entries of $\bmat{X}$ account for the cross-talk and the diagonal entries represent the transmittance and reflectance of the horizontally and vertically polarized light.
$\bmat{X}$ appears twice in the equation, where the first (right) $\bmat{X}$ represents the PBS splitting the beam into two polarization states whereas the second (right) represents the combining of the two beams.
$\bmat{R}$ represents the interference with the reference beam as $\bmat{R} = \left[H^*_{ref} \,\,\,\, 0;\,\,\, 0 \,\,\,\,  V^*_{ref}\right]$, where $H^*_{ref}$ and $V^*_{ref}$ are the complex conjugates of the horizontal and vertical components of the reference light.
$\bmat{\rho}$ is a rotation matrix accounting for the relative rotation between the polarization-delay and PD-detection modules, and $\bmat{X'}$ represents the imperfection of the PBS in the PD-detection module.
$\bmat{\eta}$ is $\left[ \eta_A \,\,\,\, 0;\,\,\, 0 \,\,\,\,\eta_B\right]$ as $\eta_A$ and $\eta_B$ are the detection efficiencies of two detectors used for PD detection.
That is to say, $\bmat{X}$, $\bmat{\rho}$, $\bmat{R}$, $\bmat{X'}$, and $\bmat{\eta}$ represent the imperfections.

The cumulative Jones matrix with these imperfections can be obtained by substituting Eq.\@ (\ref{eq:JmImperfect}) into Eq.\@ (\ref{eq:cumulativeJm}) as
\begin{equation}
	\jm{c\mathit{i}} = \bmat{\eta}\, \bmat{X'}\, \bmat{R}\, \bmat{\rho}\, \jm{out}\, \jm{r0\iti}\, \jm{out}^{-1}\, \bmat{\rho}^{-1}\, \bmat{R}^{-1}\, \bmat{X'}^{-1}\, \bmat{\eta}^{-1}.
\end{equation}
Evidently, this matrix is similar to the original cumulative Jones matrix $\jm{out}\jm{r0\iti} \jm{out}^{-1}$ shown in Eq.\@ (\ref{eq:cumulativeJm}).
A matrix and its similar matrix have the same eigenvalues, and the imperfections thus do not affect the phase retardation measurement.
Additionally, because similar self-calibration occurs for the local Jones matrix computation, the imperfections also do not affect the local phase retardation computation.
That is to say, without any special signal processing, the imperfections are automatically canceled in the derivations of the cumulative and local Jones matrices.

Although here we have discussed an example with a passive-delay based input-polarization multiplexing, the conclusion can be generalized for all types of input-polarization multiplexing.
Further details of this self-calibration are described in Section 5.4 of Ref.\@ \cite{MJJu2013OpEx} and Sections 35.5.2 and 35.5.3 of Ref.\@ \cite{Yasuno2015Book_JmOct}.

\subsection{Resolution of birefringence imaging}
\label{sec:birefResolution}
It might be fair to note that the depth resolution of birefringence imaging is lower than that of OCT.
As described in Sections \ref{sec:principle_cpr} and \ref{sec:principle_lpr}, the birefringence is computed from a local Jones matrix, and the local Jones matrix is computed from two cumulative Jones matrices at slightly distant depths.
And hence, the practical depth resolution of birefringence imaging can be roughly estimated by adding the optical axial resolution and the depth-separation of the two cumulative Jones matrices.

The separation of the cumulative Jones matrices is selected from the following three perspectives.
At first, the separation should be close enough as the tissue between the two depths can be considered as homogeneous, as discussed in Section \ref{sec:principle_lpr} of this manuscript and Sections 2, 6.1, 6.3 and Appendix A of Ref. \cite{Sugiyama2015BOE}.
Second, the separation must be larger than the optical resolution of the OCT, otherwise the two Jones matrices are highly correlated to each other and meaningful birefringence value cannot be obtained.
Third, the separation should be large enough to make the local phase retardation larger than the measurement noise of the phase retardation.

Typically, the separation is around 20 to 50 \um, while the depth resolution is a-few to ten-few micrometers.
And hence, the depth-resolution of the birefringence measurement is a-few-times lower than the depth resolution of OCT.
In our particular implementations, the optical resolution and the depth separation are, respectively, 6.2 \um and 24 \um (in tissue, for a retinal JM-OCT with a 1-\um probe wavelength) \cite{Kasaragod2017BOE} or 14.1 \um and 57.9 \um (in tissue, for a JM-OCT with 1.3-\um probe wavelength)\cite{Miyazawa2019BOE}.
And hence, the resolutions of the birefringence measurement of these systems are around 30 \um and 72 \um, which are around five-times lower than the optical resolution.

\subsection{Polarization-sensitive OCT and JM-OCT}
\label{sec:PsOct}
The JM-OCT described in this paper is a multi-contrast imaging modality based on OCT.
However, historically, JM-OCT was a variation of PS-OCT.
It is thus worth summarizing the relationship between the several PS-OCT methods and JM-OCT.

The Jones matrix analysis of PS-OCT has been demonstrated by multiple research groups including Park \etal \cite{HylePark2004OL}.
Some PS-OCT methods measure the Jones matrix or equivalent quantities but adopt other types of polarization analysis, such as Jones-Muller analysis \cite{Lippok2015OL,Villiger2016SciRep, Villiger2018Optica} and Stokes analysis \cite{WOh2008OpEx}.

A well-known PS-OCT method is the circular-polarization method, which illuminates the sample with a single state of circular polarization and measures two OCT signals using a PD detector \cite{Hee1992JOSAB, Pircher2011PRER}.
For this type of PS-OCT, the phase retardation and optic-axis orientation can be computed directly from the two detected OCT images.
Although it is not common, this type of PS-OCT hardware can be used for JM-OCT imaging.
That is to say, by assuming the unitarity of the sample, we can reconstruct a unitary Jones matrix from the two measured OCT signals, and then, we can apply the Jones matrix analyses described in Section \ref{sec:principle}.

\section{Further extensions of JM-OCT}
\label{sec:future}
In addition to the JM-OCT technologies described in this review, there are other on-going and possible future extensions of JM-OCT. 
This section gives examples of the extensions.
Additionally, open issues, which might be addressed by further development, are discussed.

\subsection{Combination with mechanical-property measurement}
The relationship between birefringence and a mechanical property of a tissue is an intriguing research topic because both can be related to collagen.
By measuring \exvivo porcine samples adopting JM-OCT and a uniaxial tension tester, Yamanari \etal and Nagase \etal showed that the polarization and mechanical properties of sclera are interrelated \cite{Yamanari2012PLoSOne, Nagase2013PLoSOne}.
Intraocular pressure and scleral birefringence have also been found to be correlated in human eyes \invivo \cite{Yamanari2014BOE}. 

One existing but not well-established JM-OCT extension is JM-OCT elastography \cite{Miyazawa2019BOE}.
By introducing a tissue-compression probe into JM-OCT, the simultaneous imaging of JM-OCT contrasts and tissue deformation was demonstrated.
By combining multiple contrasts, including tissue deformation, birefringence, and attenuation coefficients, five tissue layers of an \exvivo porcine esophagus were revealed.
Notably, these layers were not distinguishable if only one or two of the three contrasts were used.

Recently developed tissue-compression-based optical coherence elastography enables a more quantitative assessment of tissue mechanical properties (e.g., see Section 6.5 of Ref.\@ \cite{Larin2017BOE}).
It might be worth introducing such a quantitative optical coherence elastography method into JM-OCT.



\subsection{JM-OCT specific image processing}
\label{sec:imageProcessing}
Image processing methods specialized for multi-contrast JM-OCT have been demonstrated.
An early demonstration was the cumulative-phase-retardation-based segmentation of the chorio-scleral interface of the human retina \cite{Duan2012OpEx}.
This method mainly uses a single contrast of the cumulative phase retardation and is thus similar to a PS-OCT-based approach demonstrated by Torzicky \cite{Torzicky2012OpEx}.

Other examples have more intensively used multiple contrasts obtained through JM-OCT.
Myiazawa \etal first demonstrated tissue segmentation of the anterior eye using three contrasts, namely the OCT intensity, attenuation coefficient, and depth alteration ratio of cumulative phase retardation \cite{Miyazawa2009OpEx}. 
All the pixels of a JM-OCT image of the anterior eye were mapped in the three-dimensional feature space of these three contrasts, and the sclera, conjunctiva, uvea, and trabecular meshwork were segmented.
Additionally, Azuma \etal \cite{Azuma2018BOE} and Kasaragod \etal \cite{Kasaragod2018BOE} demonstrated an arithmetic computation for RPE and choroidal segmentation and machine-learning-based methods for lamina cribrosa segmentation, respectively, as mentioned in Section \ref{sec:retina}.

Miyazawa \etal demonstrated pixel grouping for JM-OCT, superpixelization \cite{Miyazawa2017BOE}.
In this method, the pixels in a cross-sectional JM-OCT image are mapped into a six-dimensional feature space of four JM-OCT contrasts and two (lateral and axial) spatial dimensions.
The four contrasts are the OCT intensity, OCTA, birefringence, and DOPU.
The pixels are clustered in this six-dimensional space such that each cluster comprises similar pixels in not only its JM-OCT contrasts (i.e., optical properties) but also its spatial proximity.
Each cluster forms a group of pixels, the superpixel, which, on the one hand, follows the morphological structure of the tissues and, on the other hand, establishes a tissue region with similar optical properties.
The superpixels can be used as kernels for contrast averaging and potentially as spatial kernels of estimators of optical properties such as birefringence \cite{Kasaragod2014OpEx, Yamanari2016BOE, Kasaragod2018BOE}.

\subsection{Other birefringence calculation methods}
\label{sec:otherBirefMethods}
As described in Sections \ref{sec:principle_lpr} and \ref{sec:noiseCorrection}, the birefringence is obtained through the local Jones matrix analysis \cite{Makita2010OpEx} and a maximum \textit{a-posteriori} estimator \cite{Kasaragod2017BOE} in our particular implementation.
On the other hand, there are some other methods to compute the birefringence (and, similarly, the local phase retardation) that are compatible with JM-OCT hardware. 

Villiger \etal used a JM-OCT hardware and a differential Mueller matrix method to compute the local phase retardation \cite{Villiger2016SciRep}.
Fan \etal combined a single-input circular-polarization PS-OCT and a Jones analysis to obtain the local phase retardation \cite{CFan2012OL}.
In addition, Stokes-based methods such as Refs.\@ \cite{Guo2004OL, PJTang2021Light} also can be implemented on JM-OCT hardware with minor modifications.

\subsection{Optic-axis-orientation measurement}
\label{sec:axisOrientation}
Lu \etal demonstrated the measurement of the orientation of the optic axis using a single-mode-fiber-based JM-OCT system \cite{ZHLu2012OL}.
Although this method successfully determined the absolute optic-axis orientation of a sample, it requires an on-site calibration.
For this calibration, two quarter-wave plates should be imaged with the sample.
This will elaborate the system or sample configuration and sacrifice the depth imaging range, and the system is thus not fully portable.
Recently, Li \etal demonstrated Jones matrix symmetrization method for optics axis measurement \cite{QYLi2018BOE, QYLi2020BOE, QYLi2020OL}.

In addition, several sophisticated methods of measuring the axis orientation have been demonstrated adopting other types of PS-OCT or JM-OCT in its widest sense.
Fan \etal demonstrated Jones-matrix-based axis-orientation analysis \cite{Fan2010IEEETBE} based on a single-polarization-input bulk PS-OCT\cite{Fan2010OpEx}, which has been adopted for PS-OCT-based tractography \cite{GYao2019XBM}.
Additionally, the measurement of the axis orientation based on Jones-Muller analysis has been demonstrated \cite{Villiger2018Optica}.
Although this method is based on Muller matrix analysis, the raw data were obtained through Jones-matrix tomography and then converted into the Muller matrix.

A Stokes-based method for axis-orientation measurement has been demonstrated \cite{PJTang2021Light, PJTang2022BOE}.
This method can also be applied to JM-OCT data by converting to the Stokes representation.

\subsection{Other polarization randomness metrics}
Although we considered only the DOPU in this paper, there are several other polarization randomness metrics that are compatible with the JM-OCT hardware.
Yamanari \etal used the spatial entropy of a set of Jones matrices obtained in a small local region as a depolarization measure \cite{Yamanari2016BOE}, and this approach has been adopted in investigating the RPE \cite{Matsuzaki2020SciRep}.
Whereas the DOPU represents the randomness of the Jones vector, i.e., the backscattered probe beam, this entropy represents the randomness of the Jones matrices.
It is thus a more direct measure of the sample's polarization randomness.
Lippok \etal used degree of polarization (DOP) to represent the polarization randomness \cite{Lippok2015OL}. 
They measured the Jones matrix distribution of the sample, converted it into a Jones-Muller matrix, and computed the DOP from the Jones-Muller matrix.
The entropy and DOP were both computed from the Jones matrix data sets, and they are thus compatible also with the narrow-sense JM-OCT.
In addition, the first demonstration of DOPU by G\"otzinger \etal in 2008 \cite{Gotzinger2008OpEx} is preceded by two similar works.
In 2007, Adie \etal used spatial variations of Stokes vectors to detect multiple scattering \cite{Adie2007OpEx}.
In 2008, Lee \etal used a quantity similar to DOPU to assess the depth-dependent alteration of scattering \cite{SWLee2008JBO}.
Both of the methods are applicable to JM-OCT.

The randomness of optic-axis orientation is also useful in highlighting tissue abnormality.
Willemse \etal defined a metric called optic-axis uniformity for the segmentation of a retinal nerve fiber layer \cite{Willemse2019OL}.
This method is based on Jones-Muller analysis and thus can be adapted to JM-OCT.

\subsection{Other open issues}
Although we independently demonstrated D-OCT and PS-OCT, we did not assess the dynamics of the polarization properties of tissues.
If we can establish a method of visualizing the temporal dynamics of the polarization properties, as preliminary demonstrated by Mukherjee \etal \cite{Mukherjee2020BISC}, we might reveal additional functional properties of the tissue.

Although we discussed noise correction methods in Section \ref{sec:noiseCorrection}, other factors may disturb the quantitative JM-OCT measurement.
One example is the defocus, which has been found to degrade the accuracy of birefringence and DOPU measurements and also generates polarization artifacts \cite{LidaZhu2022BOE}.
This effect can be mitigated by computational refocusing.

Computational adaptive optics \cite{Adie2012PNAS,Kumar2015BOE} and defocus correction \cite{Yasuno2006OpEx, Ralston2007NatPys} may further improve the image quality of OCT and JM-OCT.
However, the aberration can be polarization dependent \cite{CHe2021Arxiv}.
The combination of JM-OCT and computational aberration correction and refocusing methods may enable polarization-dependent aberration correction.

\section{Conclusion}
JM-OCT is a multi-contrast OCT modality based on a conception, in which all contrasts are derived from a single quantity, i.e., a time-sequence of Jones-matrix tomographies.
Historically, JM-OCT began as a variation of PS-OCT, but it later acquired a multi-contrast capability.
JM-OCT imaging not only relies on optical instrumentation but also exploits several mathematical and computational techniques, such as sophisticated contrast estimators.
JM-OCT has been applied to several medical and biological fields, with posterior and anterior eye imaging being two of the most successful applications.
JM-OCT has recently been applied to \invitro and \exvivo samples and \invivo small animals.
This variation of JM-OCT, JM-OCT microscopy, is expected to be a volumetric, label-free, and quantitative microscopy.
It will thus be a useful tool in biology and drug development.
Several technical issues still need to be overcome, such as the accurate optic-axis-orientation measurement and polarization-dependent computational aberration correction.
These on-going works will further enhance the capability of JM-OCT.


\section*{Acknowledgments}
The JM-OCT-related research works are based on the significant contributions of previous and current laboratory members of the Computational Optics Group at the University of Tsukuba.
In particular, seminal contributions from Masahiro Yamanari and Shuichi Makita are greatly acknowledged. 

This review also covers some research works of non-Jones matrix PS-OCT and other multi-contrast OCT imaging technologies. 
It should be acknowledged that all OCT-community members have contributed heavily to establish such multi-contrast OCT technologies.

The JM-OCT projects conducted at University Tsukuba have been supported by the Japan Science and Technology Agency through contracts with CREST (JPMJCR2105), MIRAI (JPMJMI18G8), and the Development of Advanced Measurement and Analysis Systems, and by the Japan Society for the Promotion of Science (21H01836, 22K04962, 21K09684, 18K09460, 18H01893, 17H04350, 15K13371, 15K10905, 25293353, 24592682, 22390320, 21659397).

	

	%
	%
	\bibliography{IEEEabrv,reference_yasuno2022}
	\bibliographystyle{IEEEtran}
	
	
	\begin{IEEEbiography} 
		[{\includegraphics[width=1in,height=1.25in,clip,keepaspectratio]{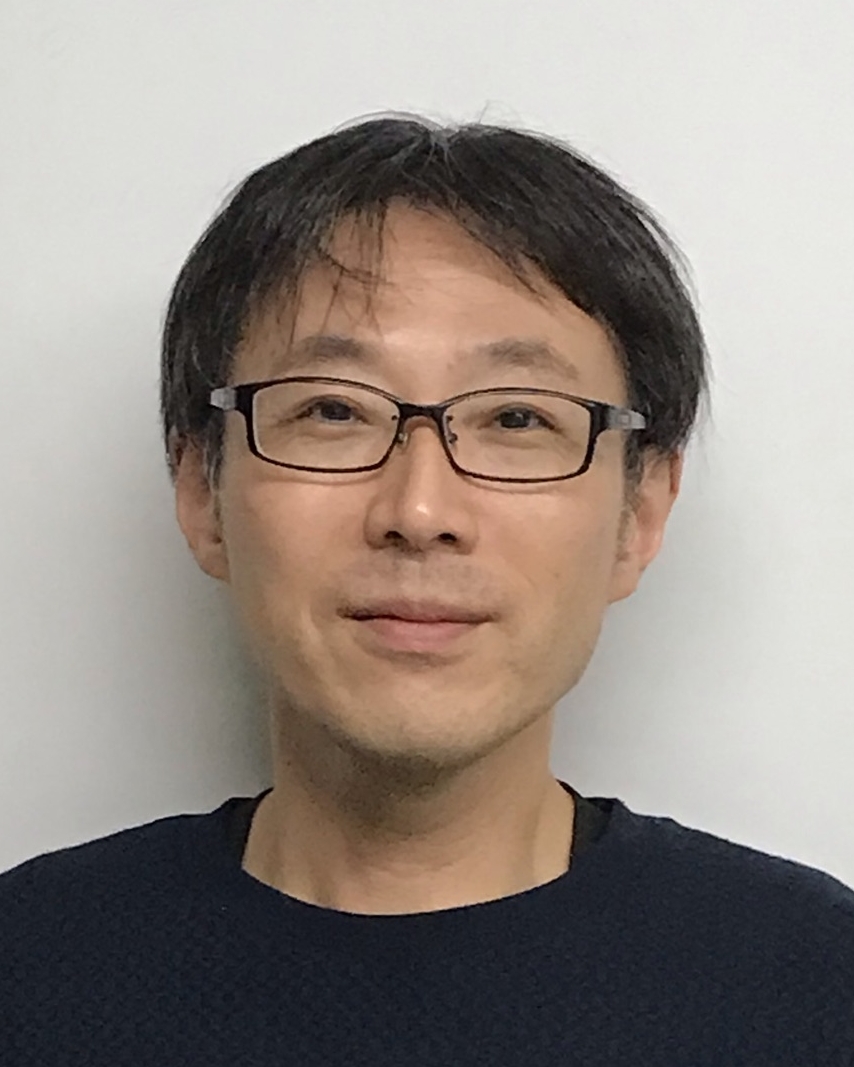}}]{Yoshiaki Yasuno}
		leads the Computational Optics Group at the University of Tsukuba.
		He received a Ph.D. degree from the University of Tsukuba for spatiotemporal optical computing in 2000 and extended this concept to optical measurements including Fourier-domain optical coherence tomography. Since 2003, he has been working on ophthalmic OCT imaging, including imaging of the retina and anterior eye.
		He has also actively worked on the contrast extension of OCT, which includes OCT angiography, polarization sensitive OCT, and JM-OCT.
		His current research interests are multi-functional optical coherence microscopy, its augmentation using computational technologies, and theoretical modeling in modern metrology.
	\end{IEEEbiography}
	
	
\end{document}